\documentclass[%
twocolumn,10pt,
 amsmath,amssymb,
 aps,prc,
]{revtex4-2}

\usepackage{graphicx}% Include figure files
\usepackage{dcolumn}% Align table columns on decimal point
\usepackage{bm}% bold math
\usepackage{float}
\usepackage[section]{placeins} %Ensure figures appear in the correct sections
\usepackage[utf8]{inputenc}
\usepackage[dvipsnames]{xcolor}
\usepackage{subfigure}
\usepackage{hyperref}
\hypersetup{
    colorlinks=true,
    citecolor=OliveGreen,
    urlcolor=cyan
}

\newcommand{\eq}[1]{Eq.~(\ref{#1})}

\usepackage{comment}

\newcommand{\pt}{p_T}

\begin{document}

\preprint{APS/123-QED}

\title{B-meson Nuclear Modification Factor and \texorpdfstring{$\mathbf{v_2(p_T)}$}{title} in a Strongly Coupled Plasma in \texorpdfstring{$\mathbf{Pb+Pb}$}{Pb+Pb} Collisions at \texorpdfstring{$\mathbf{\sqrt{s_{NN}}=2.76}$}{title} TeV and \texorpdfstring{$\mathbf{\sqrt{s_{NN}}=5.5}$}{title} TeV}

\author{B.\ A.\ Ngwenya}
\email{ngwble001@myuct.ac.za}
\author{W.\ A.\ Horowitz}%
 \email{wa.horowitz@uct.ac.za}
\affiliation{%
 Department of Physics, University of Cape Town\\
 Private Bag X3, Rondebosch 7701, South Africa
}%

\date{\today}

\begin{abstract}
We present predictions for the suppression of B-mesons using AdS/CFT techniques assuming a strongly coupled quark-gluon plasma at $\sqrt{s_{NN}}=2.76$ TeV for central collisions and $\sqrt{s_{NN}}=5.5$ TeV for various centrality classes.  We provide estimates of the systematic theoretical uncertainties due to 1) the mapping of QCD parameters to those in $\mathcal N = 4$ SYM and 2) the exact form of the momentum dependence of the diffusion coefficient predicted by AdS/CFT.  We show that coupling energy loss to flow increases $v_2$ substantially out to surprisingly large momenta, on the order of $\sim25$ GeV/c, thus pointing to a possible resolution of the $R_{AA}$ and $v_2$ puzzle for light hadrons.  
\end{abstract}
\maketitle

\section{\label{Intro}Introduction}
Heavy mesons, the decay fragments of heavy quarks, provide a uniquely powerful tool for studying the properties of the quark-gluon plasma (QGP) produced in heavy ion collisions at the Relativistic Heavy Ion Collider (RHIC) and the Large Hadron Collider (LHC).  The large mass of $b$ quarks mean they are predominantly produced during the initial nuclear overlap.  Their large mass ensures that their production processes can be described using perturbative quantum chromodynamics (pQCD) techniques \cite{Cacciari:1998it,Frixione:2001pa,Cacciari:2012ny}, which are well-tested by comparisons to the production cross sections measured in proton-proton collisions across a large range of $\surd s$ \cite{Cacciari:2005rk,Cacciari:2012ny,Lewis:2014cka}.  The large mass also implies a large scale separation between the $T\sim400$ MeV thermal QGP dynamics and that of the heavy quark itself, which leads to a well-controlled environment for theoretical predictions of energy loss processes; heavy quark energy loss may be thought of as a proving ground for energy loss techniques that may then be generalized to the smaller mass limit.

In this work we take as a starting point the assumption that the QGP created at RHIC and LHC is strongly coupled.  This assumption is based on the evidence from sophisticated viscous, relativistic hydrodynamics simulations that show that the viscosity to entropy density ratio of the QGP at RHIC and LHC is very small, $\sim0.1$ in natural units \cite{Everett:2020xug}, which is of the order predicted as a near-universal lower bound of strongly-coupled theories by AdS/CFT \cite{Kovtun:2004de,Brigante:2008gz}.  We further assume that the heavy quark probes are also strongly coupled to the QGP; i.e.\ we assume that we may model the heavy quark propagation in the QGP fully within the AdS/CFT picture \cite{Gubser:2006bz,Herzog:2006gh,CasalderreySolana:2006rq}.  For bottom quarks, this AdS/CFT picture is self-consistent up to high momenta of at least $\pt\sim100$ GeV/c \cite{Gubser:2006nz,Moerman:2016wpv}.  Qualitatively, this assumption of a heavy quark strongly coupled to a strongly coupled QGP makes intuitive sense because energy loss processes are presumably dominated by momentum transfers for which the running coupling is large; if there are large momentum transfers for which the running coupling is small, the contribution from these large momentum transfers is by definition small since the running coupling is weak in this regime.  

There are other heavy flavor energy loss models based alternative approaches, for instance pQCD calculations or an admixture of perturbative and non-perturbative physics; see \cite{Dong:2019byy} and references therein.  
These other models have at least some theoretical justification.  Perhaps future research, in particular incorporating higher order contributions, will provide insight into the $p_T$ and $T$ ranges of applicability of the various heavy quark energy loss models.  In the meantime, we take a phenomenological approach in which we seek to understand whether or not the assumption of a strongly coupled plasma, strongly coupled to $b$ quarks is at least qualitatively consistent with current experimental data.  To this end, in this work we demonstrate a consistency between our predictions of heavy meson suppression at $\surd s_{NN} = 2.76$ TeV with our own previous predictions \cite{Horowitz:2015dta} and measurements from LHC \cite{Khachatryan:2016ypw,Sirunyan:2017oug}.  We then provide predictions for suppression at $\surd s_{NN} = 5.5$ TeV for comparison with future LHC measurements.

\section{Energy Loss Model}
\begin{figure}[!htbp]
\centering
\includegraphics[width=0.9\linewidth]{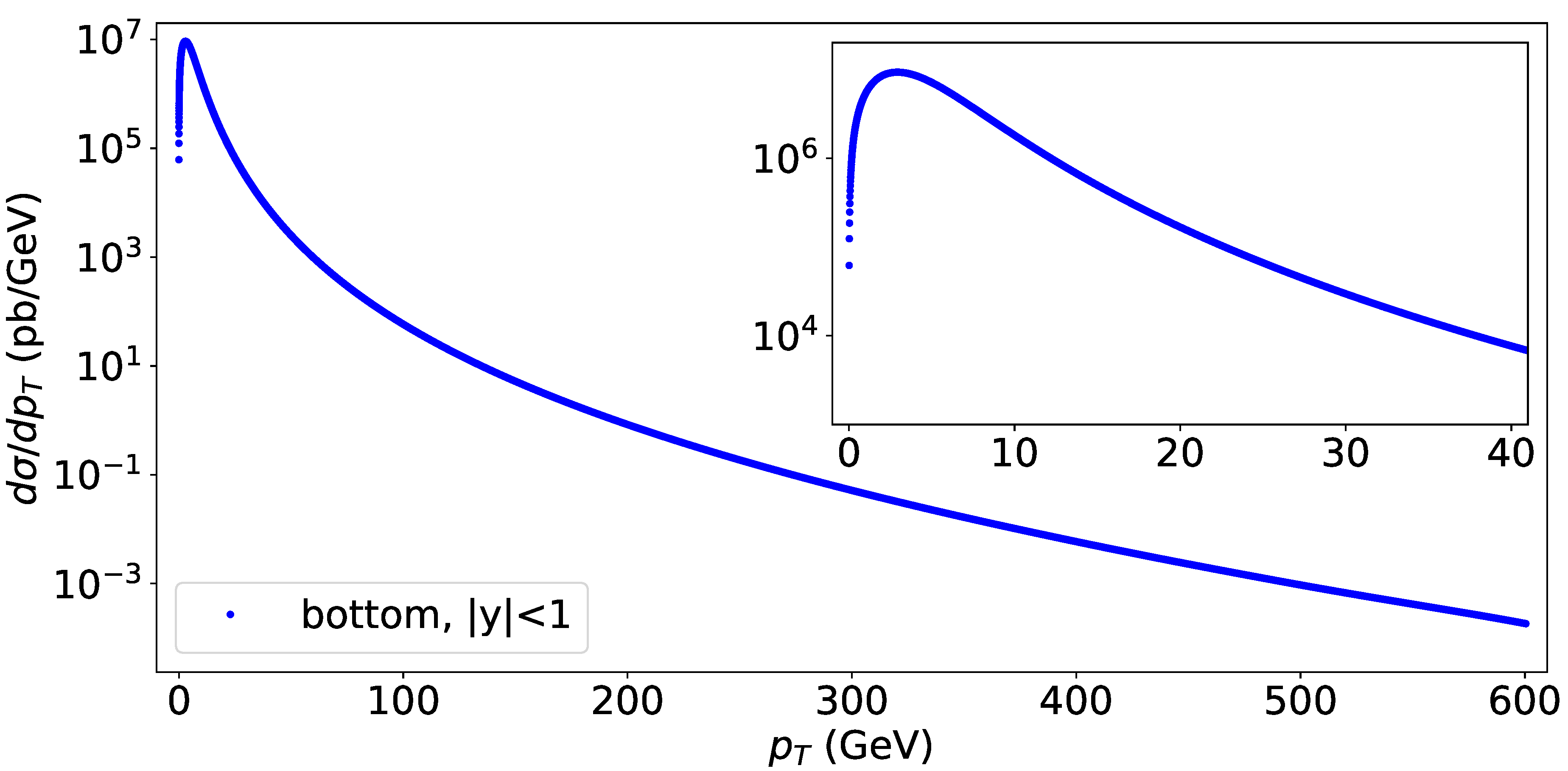}
\caption{\label{Fig1} Bottom quark production in pp collisions at the LHC at $\sqrt{s_{NN}}=5.5$ TeV for $|y|<1$, the $y$-axis is in log-scale.}
\end{figure}

Our work is based on the energy loss model developed in \cite{Horowitz:2015dta} and expanded on in \cite{Hambrock:2017sno,Hambrock:2018sim}. In particular, the production spectrum of the bottom quarks is obtained from fixed order, next to leading log (FONLL) calculations \cite{Cacciari:1998it,Frixione:2001pa,Cacciari:2012ny} and is shown in Fig.\ \ref{Fig1} for $\surd s_{NN} = 5.5$ TeV for $|y|<1$. The spectrum peaks at $p_T \sim 3$ GeV/c and has a power law dependence on momentum. 

We assume that the heavy quarks fragment in vacuum, and hence we may employ the FONLL fragmentation functions \cite{Cacciari:2005uk}.  The FONLL B meson fragmentation functions are based on the Kartvelishvili et al.\ distribution \cite{Kartvelishvili:1977pi},
\label{Eq 7}
\begin{eqnarray}
D_{NP}^{b \rightarrow B} (z) =&& (\alpha +1)(\alpha +2)z^{\alpha}(1-z), \nonumber \\
&&z=\frac{p_{meson}}{p_{quark}},
\end{eqnarray}
with $\alpha=24.2$ and $m_b=4.75$ GeV \cite{Cacciari:2005uk}.  

We assume that the heavy quarks are produced at $t=0$ fm/c, are distributed in the transverse plane according to the binary collision density given by the Optical Glauber model \cite{Miller:2007ri}, and are initially distributed uniformly in angle about the reaction plane. We used a Woods-Saxon density distribution for the nuclei
\begin{eqnarray}
\rho (r)    &=&     \frac{\rho _0}{1+ exp\left[ \frac{r-R}{a} \right]}
\end{eqnarray}
with $R=6.624$ fm and $a=0.549$ fm \cite{Loizides:2017ack}.  Further, we took $\sigma_{inel}^{NN}=6.85$ fm$^2$ at $\surd s_{NN} = 5.5$ TeV \cite{Loizides:2017ack}. 

For the medium evolution, we used the results from VISHNU 2+1D viscous relativistic hydrodynamics \cite{Shen:2011eg,Qiu:2011hf}.

For the energy loss of the heavy quark strongly coupled to a strongly coupled plasma, we used results from the AdS/CFT correspondence.  The original AdS/CFT calculations of energy loss computed only the mean energy loss, or drag, \cite{Gubser:2006bz,Herzog:2006gh} or the diffusion coefficient at non-relativistic speeds \cite{CasalderreySolana:2006rq}.  These early calculations and the first phenomenological energy loss model \cite{Akamatsu:2008ge} assumed that the fluctuation-dissipation theorem held for heavy quark energy loss.  Gubser \cite{Gubser:2006nz} explicitly computed the diffusion for the hanging, dragging string setup of \cite{Gubser:2006bz,Herzog:2006gh}, in which the heavy probe quark is dragged for all time at a constant velocity.  In this constant quark velocity setup, Gubser found that the diffusion grew rapidly, $\sim\gamma^{5/2}$, in the direction parallel to heavy quark motion, and that the diffusion was not related to the drag by the Einstein relations except in the limit $v\rightarrow0$.  
Subsequent AdS/CFT calculations of both heavy and off-shell light quark diffusion \cite{Moerman:2016wpv} suggest that the growth in the diffusion coefficient seen by Gubser is likely an artifact of the setup in his calculation that the speed of the heavy quark is held constant (by an external force constantly pumping in energy and momentum); this subsequent work also suggests that for the phenomenologically relevant case of heavy quarks undergoing unforced motion, the diffusion coefficient is in fact momentum independent.

Thus, in an effort to capture a sense of the systematic uncertainty in phenomenological B meson suppression predictions from the current uncertainty in the precise momentum dependence of the drag and diffusion coefficients predicted by AdS/CFT, we employ two assumptions about the momentum dependence of the drag and the diffusion.  The first scenario, which we denote $\mathbf{D(p)}$, assumes the diffusion coefficient grows rapidly with $\gamma$, and is described in detail in \cite{Horowitz:2015dta}.  The second scenario, which we denote $\mathbf{D=const}$, assumes the diffusion coefficient does not depend on the speed of the heavy quark, with the drag given by the Einstein relations, and is described in \cite{Hambrock:2018sim}.

In detail, and as described in \cite{Horowitz:2015dta}, in the D(p) scenario, the Stratonovich stochastic differential equation implemented as an It\^o SDE in the Euler-Maruyama scheme is
\begin{align}
	p'^i_{n+1} 
	& = \Big[1 - \mu \, dt' + \frac{1}{2} \kappa \, dt' \Big( \, \frac{5\gamma^{5/2}}{4E'^2} \nonumber \\ 
	& \qquad + \frac{(d-1) \, \gamma^{1/2}}{(\gamma^2+1) \, M_Q^2} \, \Big)  \Big] p'^i_n + C^{ij} dW_j \\
	\mu & = \frac{\pi\sqrt{\lambda}T^2}{2M_Q} \\
	C^{ij} & = \sqrt{dt'\,\kappa}\gamma^{1/4}\left( \frac{p'^ip'^j}{(\gamma^2+1) \, M_Q^2} + \delta^{ij}\right).
\end{align}
where $M_Q$ is the mass of the heavy quark, $dt'$ is the time step $dt$ boosted into the local rest frame of the fluid; $dt' \, = \, dt/\gamma$; $\kappa \, = \, \pi\sqrt{\lambda}T^3$, where $T$ is the temperature of the fluid in its local rest frame; $d$ is the number of spatial dimensions in the calculation (in this case, we propagate the heavy quarks through backgrounds generated by VISHNU \cite{Shen:2011eg,Qiu:2011hf}, which is a $2+1D$ hydrodynamics code); the $dW_j$ are the uncorrelated, Gaussian Wiener kicks with mean zero and standard deviation one.

In detail, in the D=const scenario, the It\^o SDE in the Euler-Maruyama scheme we implement is
\begin{align}
	p'^i_{n+1} & = \Big(1 - \mu \, dt' \Big) p'^i_n + C^{ij} dW_j \\
	\label{eq:dragconst}
	\mu & = \frac{\pi\sqrt{\lambda}T^2}{2E'} \\
	C^{ij} & = \sqrt{2dt'\,\mu E'\,T} \delta^{ij}.
\end{align}

In both the D(p) and D=const scenarios, the heavy quark is propagated in coordinate space according to
\begin{align}
	x^i_{n+1} = \frac{p^i_{n+1}}{E_{n+1}}dt,
\end{align}
$dt$ was taken to be $1/150\times\mu_{max}$, where $\mu_{max}$ is the drag coefficient at the center of the fireball at the thermalization time, the largest drag coefficient for any individual collision \cite{Horowitz:2015dta}.  

\begin{figure}[!htbp]
\centering
\includegraphics[width=0.9\linewidth]{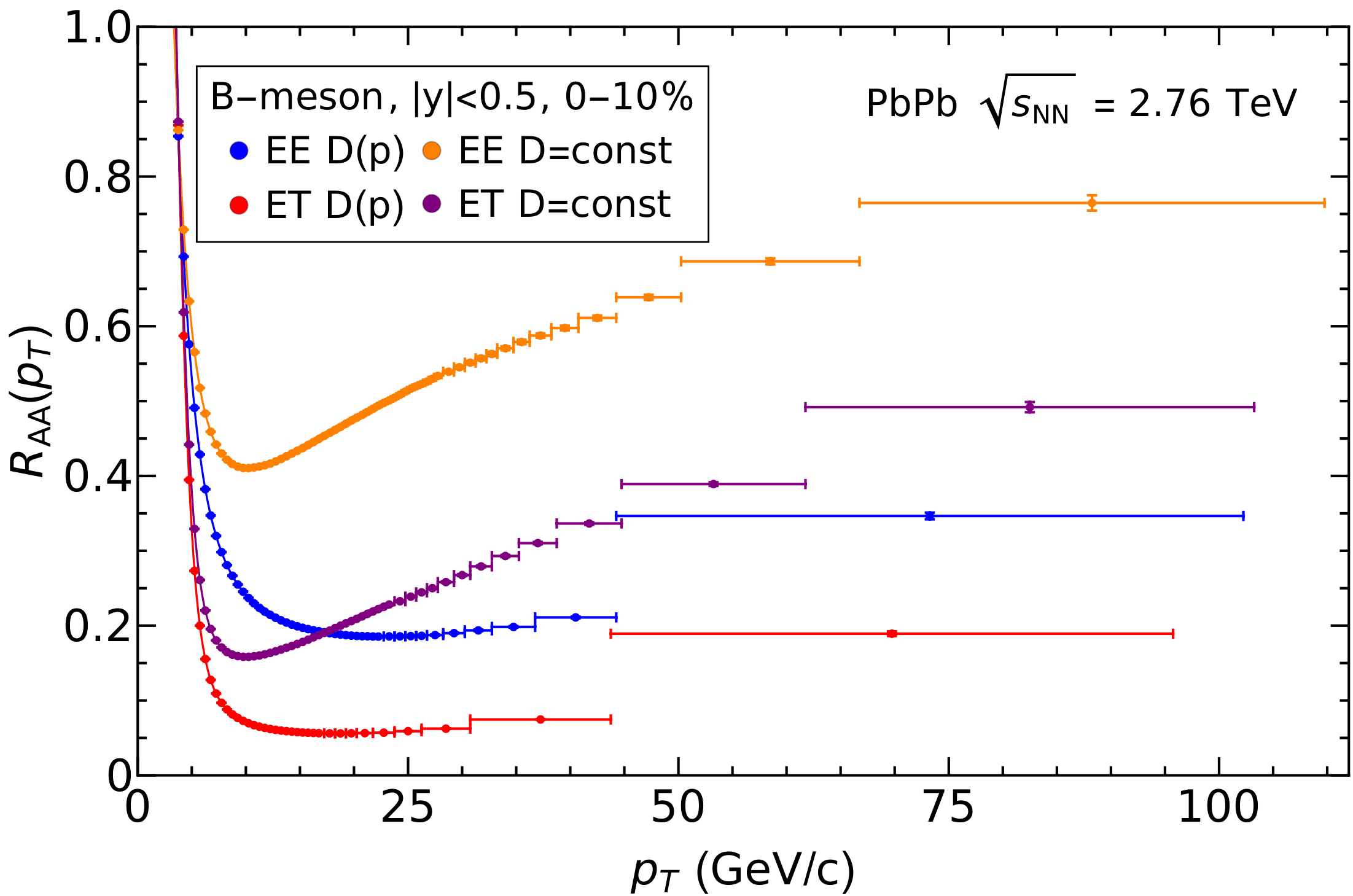}
\caption{\label{Fig2}B-meson $R_{AA}(p_T)$ at $\sqrt{s_{NN}} =2.76$ TeV, 0-10\% centrality for EE and ET parameters with a constant diffusion coefficient and one that is dependent on the momentum.}
\end{figure}
\begin{figure}[!htbp]
\centering
\includegraphics[width=0.9\linewidth]{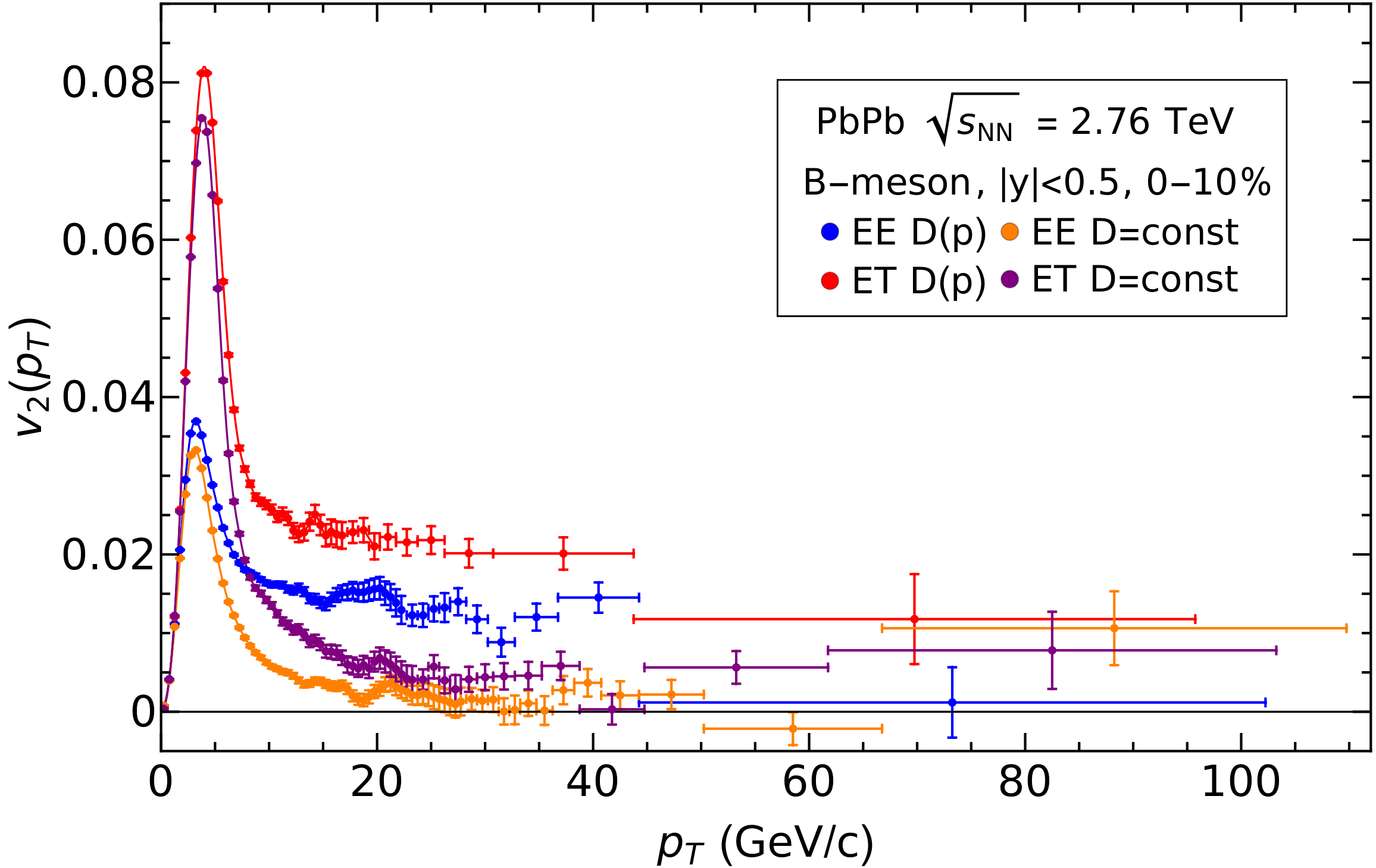}
\caption{\label{Fig3}$v_{2}(p_T)$ for B-mesons at $\sqrt{s_{NN}} =2.76$ TeV with diffusion coefficients, D=const as well as D(p), for both EE and ET parameters at 0-10\% centrality.}
\end{figure}

The drag and diffusion coefficients have temperature and 't Hooft coupling dependencies that we must fix in our energy loss model.  The mapping between the phenomenologically relevant physical parameters from QCD to the parameters in $\mathcal{N}=4$ super Yang-Mills (SYM) is not completely fixed.  We've used two different mappings between the QCD and $\mathcal N = 4$ SYM parameters in order to (at least partially) explore the systematic theoretical uncertainty: 
\label{Eq 22}
\label{Eq 23}
\begin{enumerate}
    \item \textbf{Equal Temperature and Parameters (ET):}
    \begin{equation}
        T_{SYM}=T_{QCD}, \quad \lambda =  4\pi \alpha_s N_c = 4\pi \times 0.3 \times 3 \simeq 11.3
    \end{equation}
    \item \textbf{Equal Energy Density and HQ Potential (EE):}
    \begin{equation}
        T_{SYM}=\frac{1}{3^{1/4}}T_{QCD}, \quad \lambda = 5.5
    \end{equation}
\end{enumerate}
The Equal Temperature and Parameters assume the temperature of the QCD plasma is the same as the temperature of the $\mathcal{N}=4$ SYM plasma.  Further, the 't Hooft coupling is fixed by equating the coupling in $\mathcal{N}=4$ SYM to the coupling in QCD, $g_{YM}=g_s$.   
The Equal Energy Density and HQ Potential framework assumes the energy density of the QCD plasma is the same as the $\mathcal N = 4$ SYM plasma, $\epsilon_{SYM} =\epsilon_{QCD}$; the overall factor of $1/3^{1/4}$ between $T_{QCD}$ and $T_{SYM}$ is then due to the approximately 3 times greater number of degrees of freedom in $\mathcal N=4$ SYM compared to QCD \cite{Gubser:2006qh}.  Further, the 't Hooft coupling in $\mathcal{N}=4$ SYM is determined by comparing the static force between a quark and antiquark, yielding $\lambda = 5.5$ \cite{Gubser:2006qh}. Note that in previous proceedings of similar work, the ET and EE parameters were referred to as ``$\alpha_s=0.3$'' and ``$\lambda=5.5$,''  respectively, in \cite{Horowitz:2015dta} and ``Reasonable'' and ``Gubser,'' respectively, in \cite{Hambrock:2017sno,Hambrock:2018sim}.

\begin{figure}[!tbp]
    \centering
    \subfigure[\label{Fig4a}]{{\includegraphics[width=0.9\linewidth]{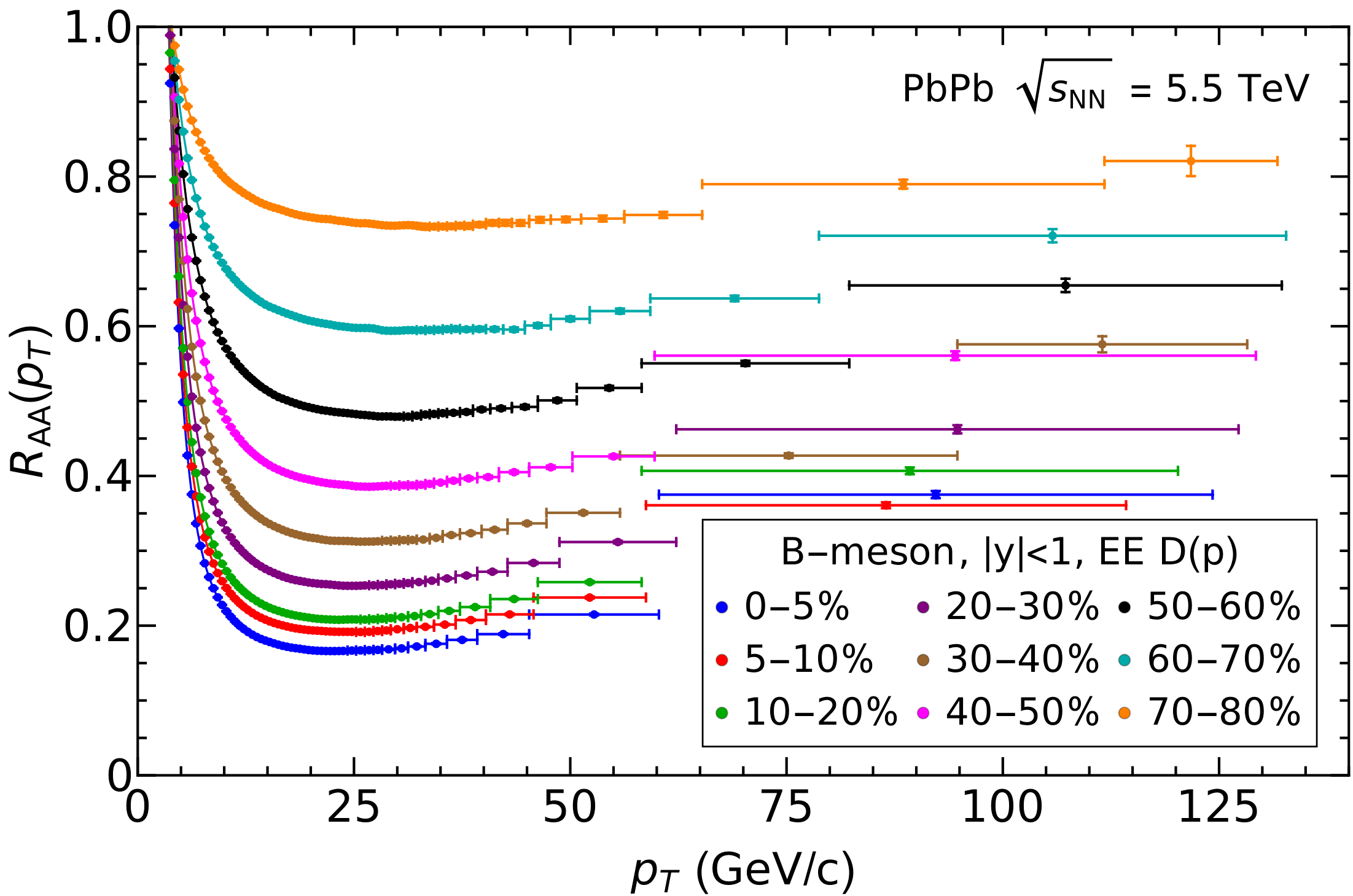} }}%
    \hspace{1em}
    \subfigure[\label{Fig4b}]{{\includegraphics[width=0.9\linewidth]{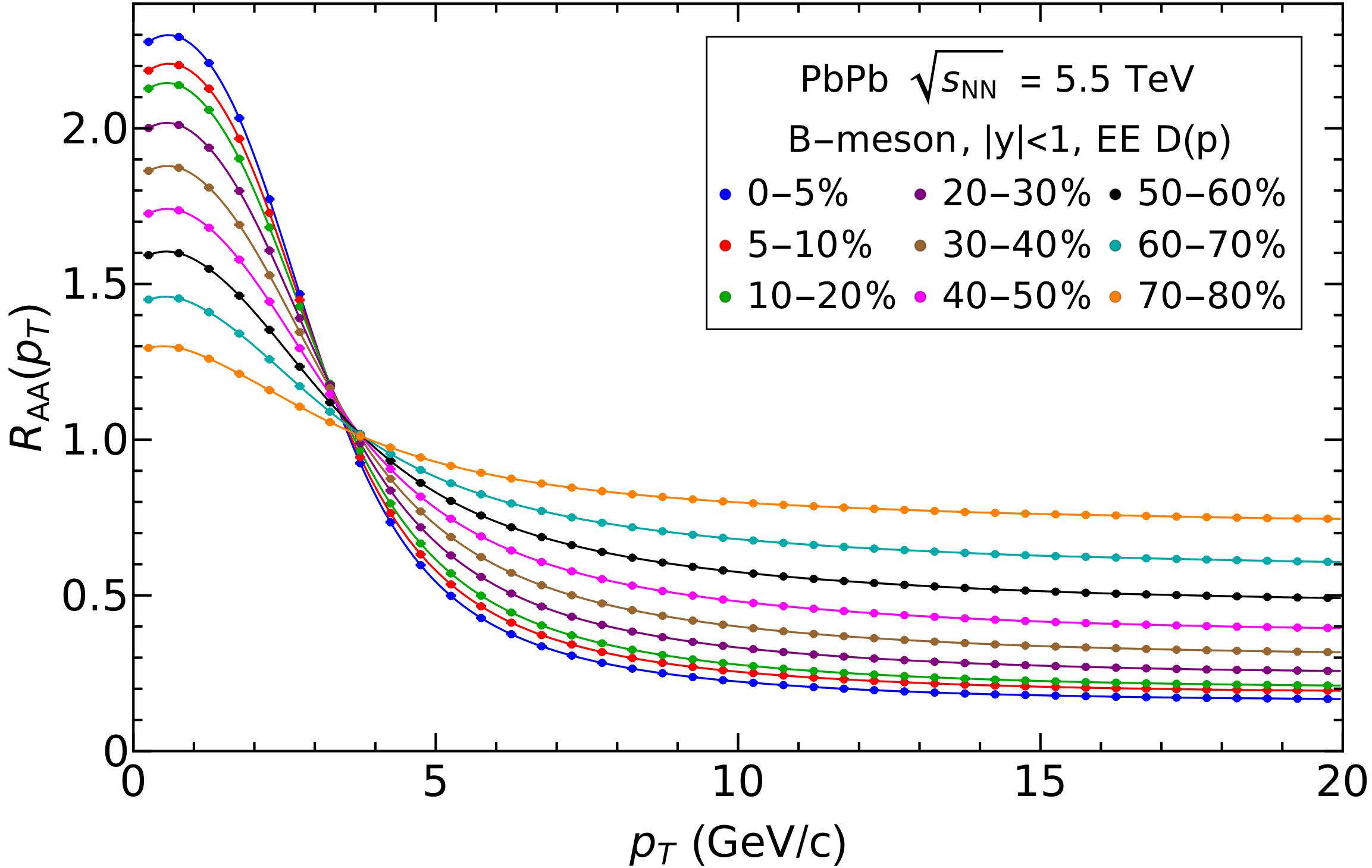} }}%
    \caption{\label{Fig4}(\hyperref[Fig4a]{a}) EE, D(p) B-meson $R_{AA}(p_T)$ at $\sqrt{s_{NN}} =5.5$ TeV for centrality classes 0-5\% up to 70-80\%. (\hyperref[Fig4b]{b}) Expanded view of the transverse momentum region, $0<p_T \leq 20$ GeV/c of (a), including the region $R_{AA}(p_T)>1$.}%
\end{figure}

In this paper we'll explore the four different combinations (setups) of the two diffusion coefficient momentum dependence scenarios and the two different parameter mappings. These four different setups give a sense of the systematic theoretical uncertainties associated with making nuclear modification factor predictions using an AdS/CFT energy loss model.

\section{\label{results}Results}

The main results of this paper are the nuclear modification factor, $R_{AA}(p_T)$ and the $v_2(p_T)$ for B mesons at $\surd s_{NN} = 2.76$ TeV, $|y|<0.5$ for central $Pb+Pb$ collisions and $\surd s_{NN} = 5.5$ TeV, $|y|<1$ for $Pb+Pb$ collisions at various centralities.  Note that the rapidity range only refers to the production cross-section; the energy loss calculation is performed in two spatial dimensions and is assumed to be at midrapidity.  

Results were initially binned in bins of width of 0.5 GeV/c.  We then set a threshold of 8500 B mesons per bin.  Bins that didn't satisfy this threshold were combined until the threshold was reached, except for some of the highest-$p_T$ bins for which we ran out of statistics.  

In all the plots that follow, we provide an interpolation line for bins of $p_T\le20$ GeV/c to help guide the eye.  One can see that the statistics are so high for these bins that there's no statistical fluctuations and such an interpolation is extremely well justified.  

\subsection{\label{276} \texorpdfstring{$\mathbf{R_{AA}(p_T)}$}{276} and \texorpdfstring{$\mathbf{v_{2}(p_T)}$}{276} at \texorpdfstring{$\mathbf{\sqrt{s_{NN}} =2.76}$}{276} TeV}

In Fig.\ \ref{Fig2} we show the results of our energy loss model for $R_{AA}(p_T)$ for B-mesons at $\surd s_{NN} = 2.76$ TeV at $0-10\%$ centrality in the rapidity range $|y|<0.5$.  We've provided these calculations in order to make contact with our previous work \cite{Horowitz:2015dta} and also recent measurements from the LHC.  The EE D(p) and ET D(p) results here are in exact agreement with our previous calculations; the D=const results are new.  The D(p) $R_{AA}(p_T)$ increases faster with $p_T$ than the D=const $R_{AA}(p_T)$, which is expected as greater fluctuations lead to smaller suppression.  Our results are qualitatively consistent with CMS non-prompt $J/\Psi$ results obtained in Ref.\ \cite{Khachatryan:2016ypw}.  However, this qualitative comparison has some limitations since the non-prompt $J/\Psi$ only takes a fraction of the B-meson's momentum in the decay $B \rightarrow J/\Psi$; further, the CMS result covers the centrality range 0-100\% and rapidity range $|y|<2.4$.  It's known experimentally and understood theoretically that the nuclear modification factor decreases with rapidity \cite{Chen:2015iga,Abelev:2013ila} and increases with centrality.  
It's important to mention that previous EE and ET results for both the D(p) and D=const schemes \cite{Hambrock:2017sno} also showed a quantitative agreement with B meson suppression as measured by CMS at $\surd s_{NN} = 5.02$ TeV \cite{Sirunyan:2017oug}.

%%%%%%%%%%%%%%%%%%%%%%%%%%%%%%%%%%%%%%%%%%%%%%%%%%%%%%%%%%%%%%%%%%%%%%%%%%%%%%%%%%%%%%%%%%%%%%%%%%%%%%%%%%%%%%%%%%%%%%

\begin{figure}[!htbp]
    \centering
    \subfigure[\label{Fig5a}]{{\includegraphics[width=0.9\linewidth]{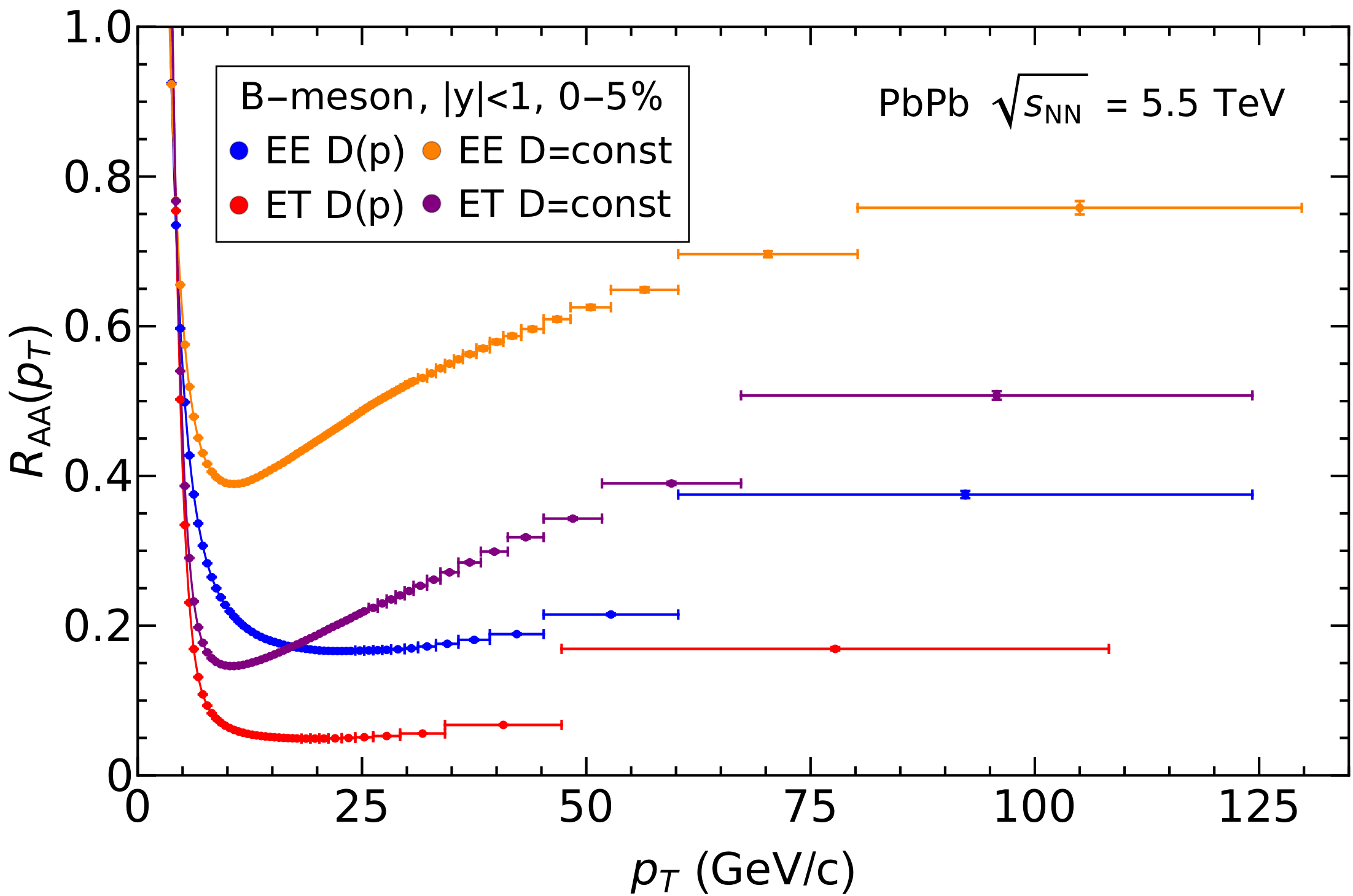} }}%
    \hspace{1em}
    \subfigure[\label{Fig5b}]{{\includegraphics[width=0.9\linewidth]{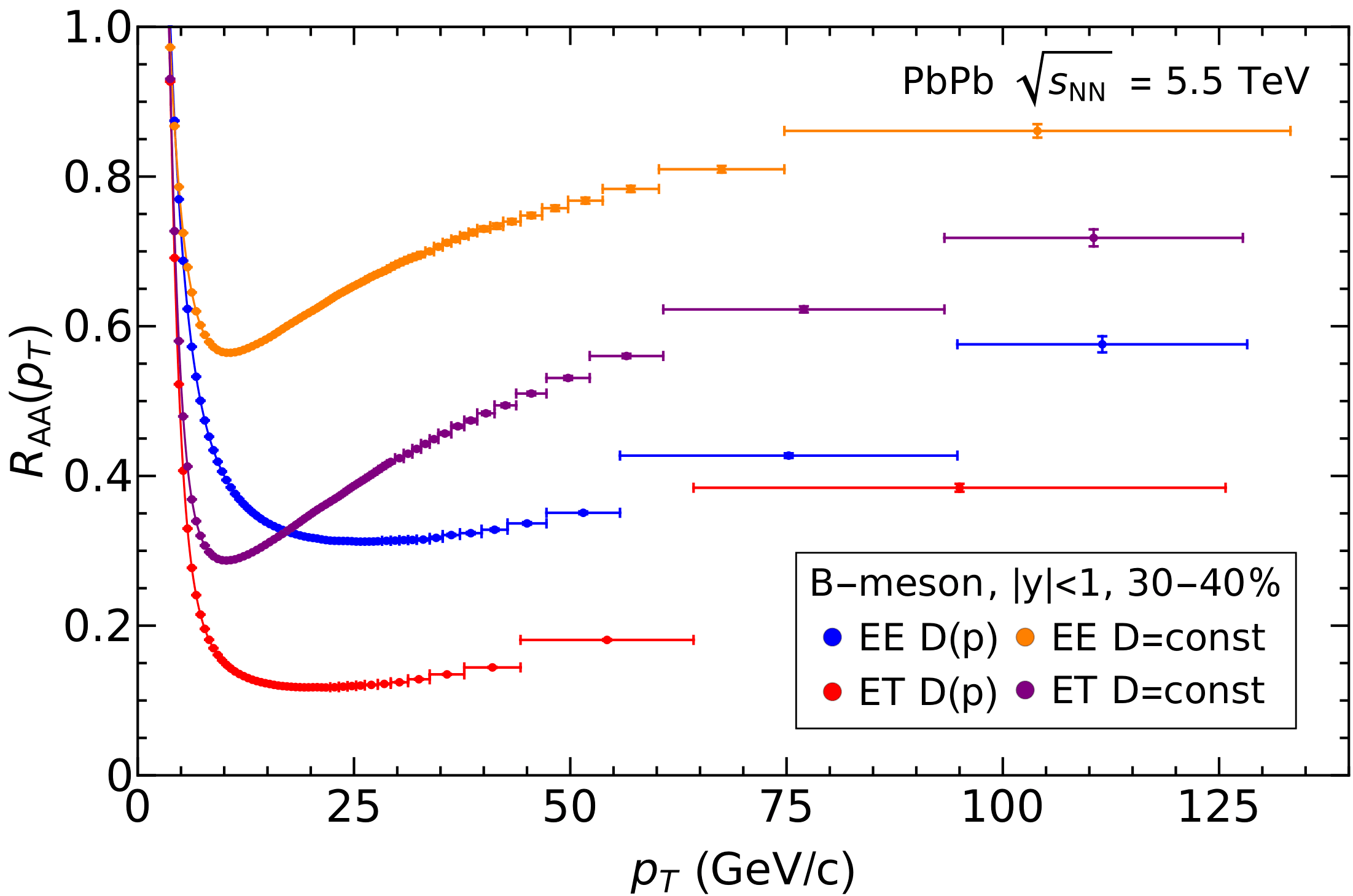} }}%
    \hspace{1em}
    \subfigure[\label{Fig5c}]{{\includegraphics[width=0.9\linewidth]{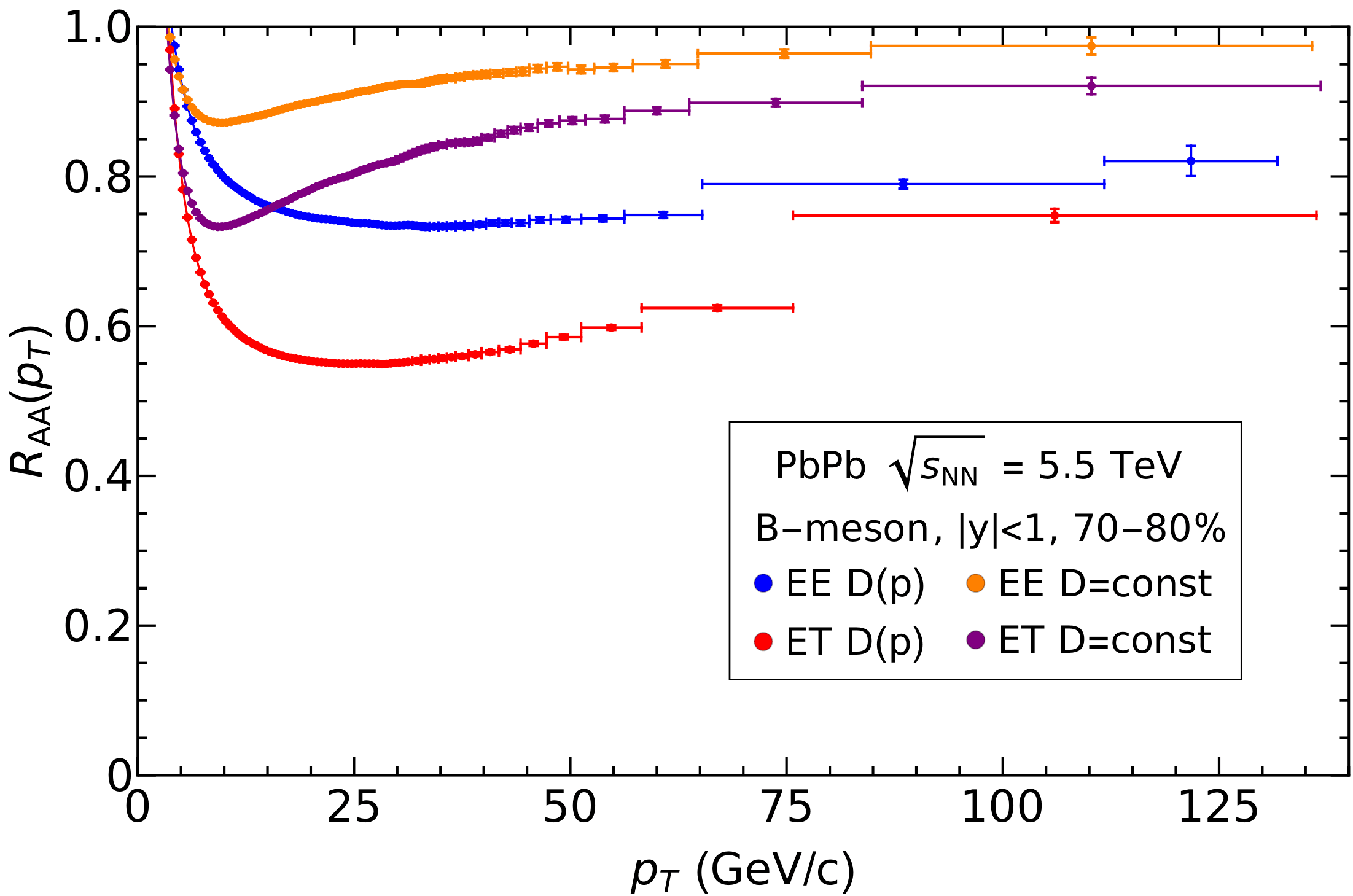} }}%
    \caption{\label{Fig5}B-meson $R_{AA}(p_T)$ at $\sqrt{s_{NN}} =5.5$ TeV for the centrality classes (\hyperref[Fig5a]{a}) 0-5\%, (\hyperref[Fig5b]{b}) 30-40\% and (\hyperref[Fig5c]{c}) 70-80\%.}%
\end{figure}

In Fig.\ \ref{Fig3} we show the results of our energy loss model for the $v_{2}(p_T)$ for B-mesons at $\surd s_{NN} = 2.76$ TeV at $0-10\%$ centrality in the rapidity range $|y|<0.5$.  Again, the results shown here for the EE D(p) and ET D(p) models are consistent with those shown previously \cite{Horowitz:2015dta}.  Note however that the statistical uncertainties quoted here correct a mistake in the uncertainty estimations made in \cite{Horowitz:2015dta}.  
These predictions are also qualitatively consistent with CMS non-prompt $J/\Psi$ data obtained in Ref.\ \cite{Khachatryan:2016ypw}, with the same limitations as for $R_{AA}(p_T)$.  In particular, the CMS $v_2(p_T)$ is slightly higher than the $v_2(p_T)$ from our calculations since $v_2(p_T)$ increases in semi-central collisions \cite{Aad:2014eoa}.  However, $v_2$ doesn't have a strong rapidity dependence \cite{Aad:2014eoa,Aaboud:2018ttm}.

\subsection{\label{RAA} \texorpdfstring{$\mathbf{R_{AA}(p_T)}$}{RAA} at \texorpdfstring{$\mathbf{\sqrt{s_{NN}} =5.5}$}{RAA} TeV}

We now present the nuclear modification factor results for B mesons at $\surd s_{NN} = 5.5$ TeV, starting with the centrality dependence of $R_{AA}(p_T)$ for EE parameters with a diffusion coefficient that is dependent on momentum shown in Fig.\ \ref{Fig4}.  In Fig.\ \hyperref[Fig4]{4} (\hyperref[Fig4a]{a}), we show the full momentum range of our results, limited only by statistics.  In Fig.\ \hyperref[Fig4]{4} (\hyperref[Fig4b]{b}) we zoom in the results for $0 < p_T \le 20$ GeV/c for clarity.  As expected, there is less suppression as one moves from central to peripheral collisions. The centrality dependence of $R_{AA}(p_T)$ plots for the rest of the other three setups are given in Appendix \ref{plots}.

%%%%%%%%%%%%%%%%%%%%%%%%%%%%%%%%%%%%%%%%%%%%%%%%%%%%%%%%%%%%%%%%%%%%%%%%%%%%%%%%%%%%%%%%%%%%%%%%%%%%%%%%%%%%%%%%%%%%%%

\begin{figure}[!tbp]
    \centering
    \subfigure[\label{Fig6a}]{{\includegraphics[width=0.9\linewidth]{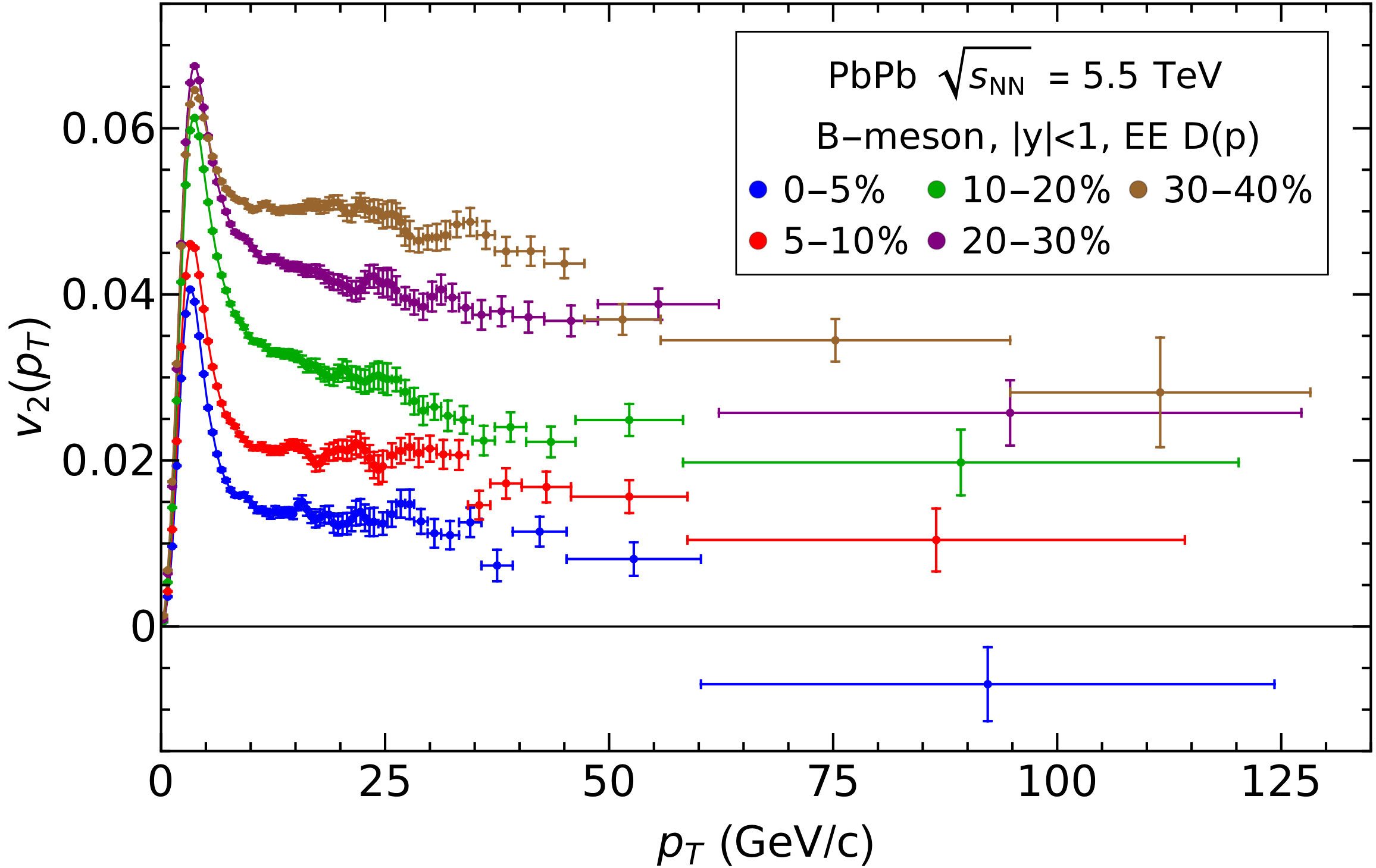} }}%
    \hspace{1em}
    \subfigure[\label{Fig6b}]{{\includegraphics[width=0.9\linewidth]{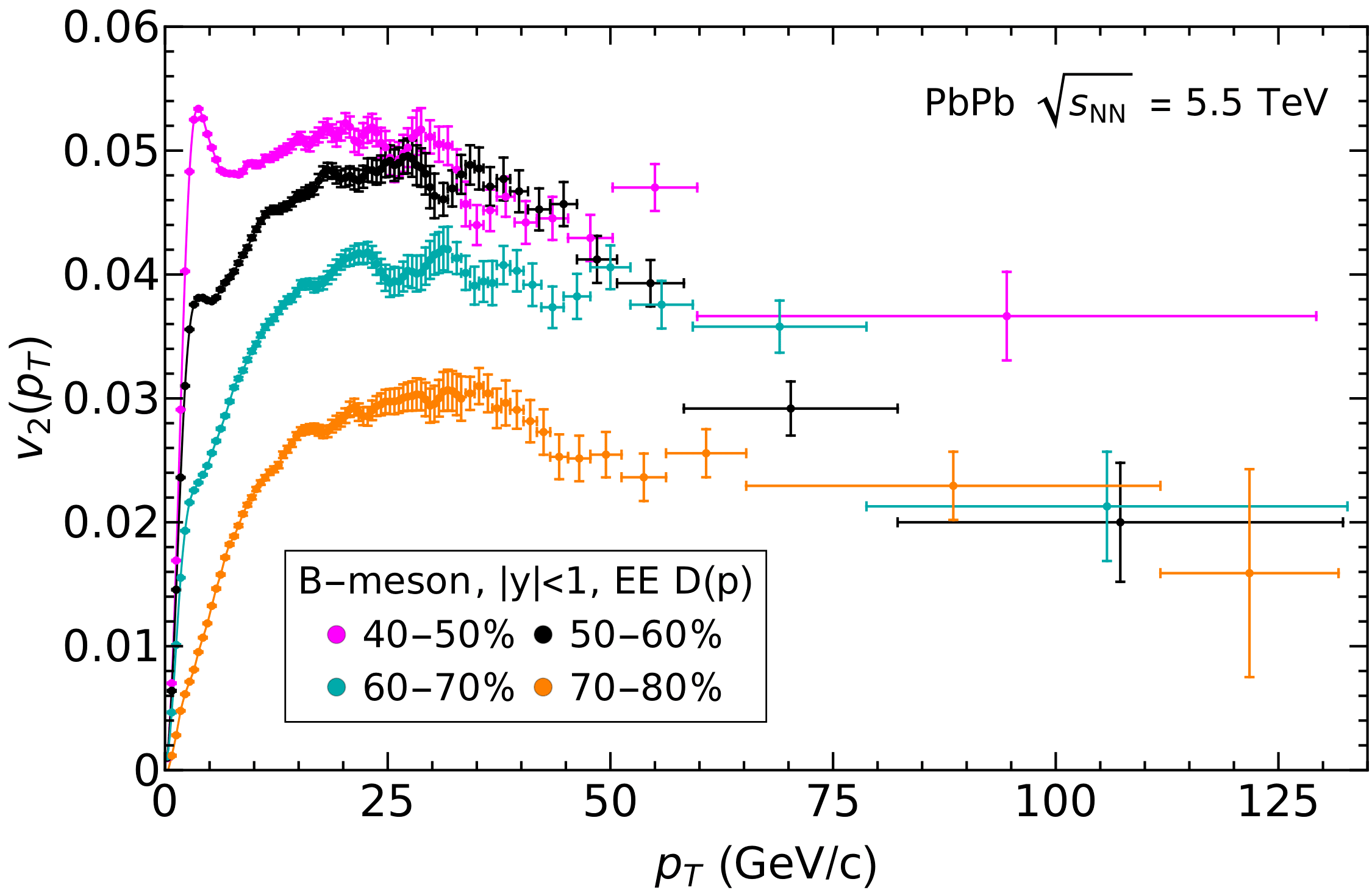} }}%
    \caption{\label{Fig6}EE D(p) B-meson $v_{2}(p_T)$ at $\sqrt{s_{NN}} =5.5$ TeV for the centrality classes (\hyperref[Fig6a]{a}) 0-5\% up to 30-40\% and (\hyperref[Fig6b]{b}) 40-50\% up to 70-80\%.}%
\end{figure}

One can also examine the nuclear modification factor per centrality class to see how $R_{AA}(p_T)$ varies for the different setups we've employed.  (Recall that the four setups come from the two different assumptions for the momentum dependence of the heavy quark diffusion coefficient and from the two different mappings from QCD parameters to $\mathcal N=4$ SYM parameters.)  These results are shown in Fig.\ \ref{Fig5} for central, semi-central, and peripheral collisions. 

Our energy loss model predicts in general that $R_{AA}(p_T)$ decreases very slightly with  $\surd s_{NN}$ from 2.76 TeV to 5.5 TeV.  This  $\surd s_{NN}$ dependence is non-trivial.  As  $\surd s_{NN}$ increases, the production spectrum of heavy quarks hardens, reducing the effectiveness of the energy loss: for a given fixed energy loss, $R_{AA}(p_T)$ is larger for harder production spectra \cite{Horowitz:2012cf}.  However, as  $\surd s_{NN}$ increases, the medium temperature and lifetime of the QGP also increase, which leads in general to a greater amount of energy lost.  The strong temperature dependence of the energy loss predicted by AdS/CFT compensates the reduction in effectiveness of the energy loss due to the hardening of the production spectrum.

The systematic difference between the $R_{AA}(p_T)$ predicted by the ET and the EE parameter mappings can be easily understood.  The drag coefficient $\mu$ depends linearly on the 't Hooft coupling $\lambda$ and quadratically on the temperature $T$.  The EE parameter mapping yields an 't Hooft coupling approximately a factor of two smaller than the ET mapping, and the temperature in the EE mapping is also smaller than for the ET mapping.

\begin{figure}[!tbp]
    \centering
    \subfigure[\label{Fig7a}]{{\includegraphics[width=0.9\linewidth]{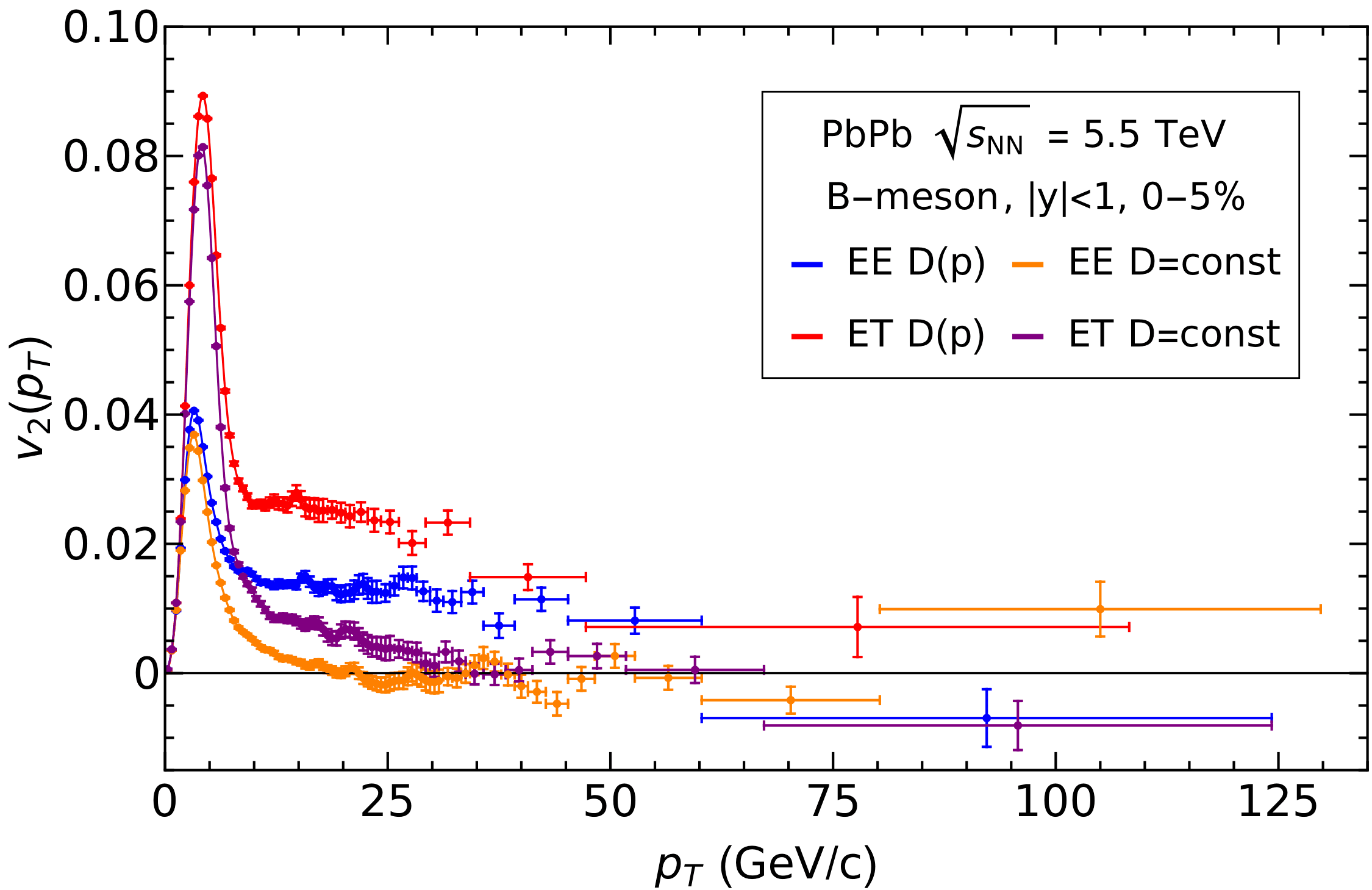} }}%
    \hspace{1em}
    \subfigure[\label{Fig7b}]{{\includegraphics[width=0.9\linewidth]{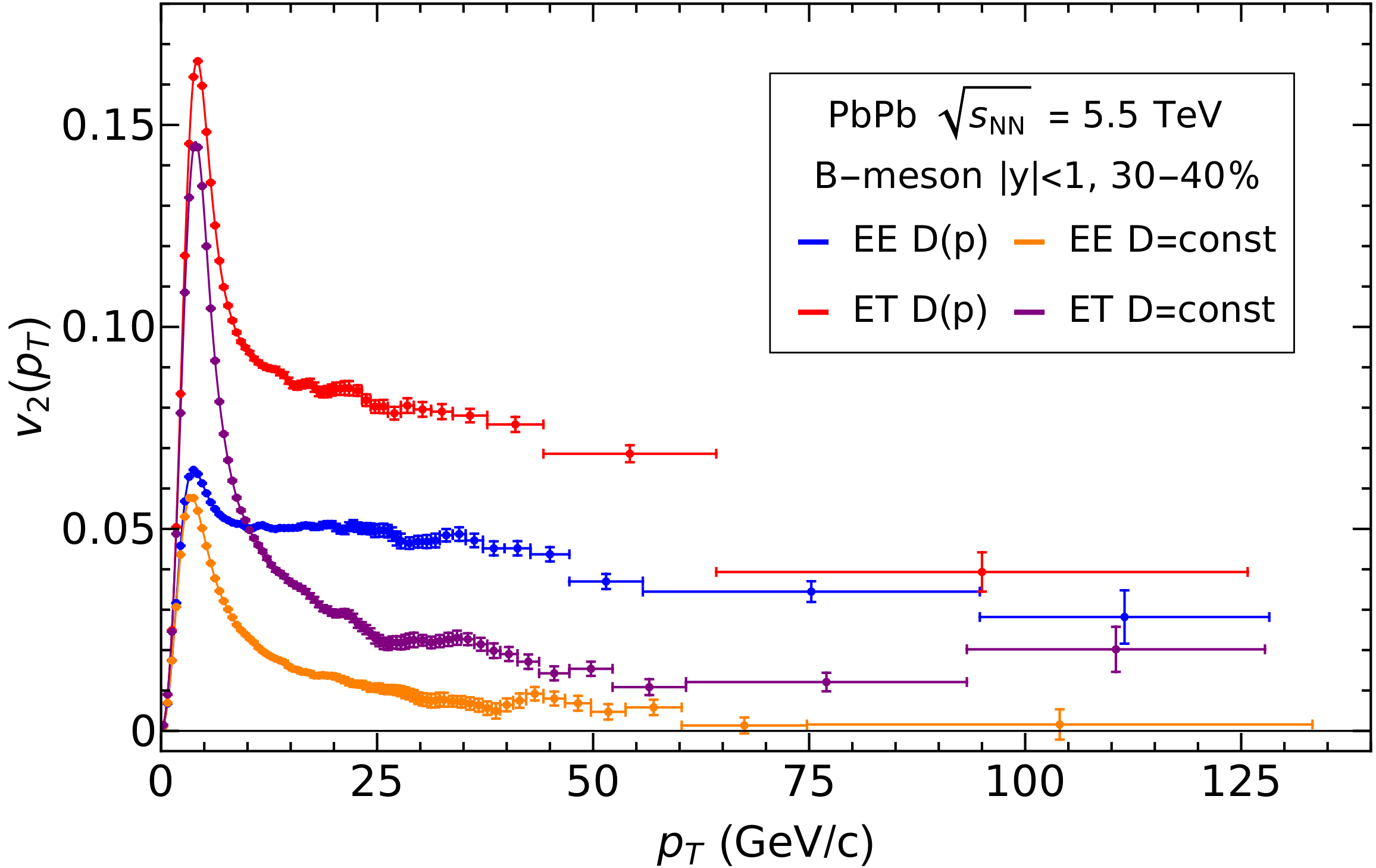} }}%
    \hspace{1em}
    \subfigure[\label{Fig7c}]{{\includegraphics[width=0.9\linewidth]{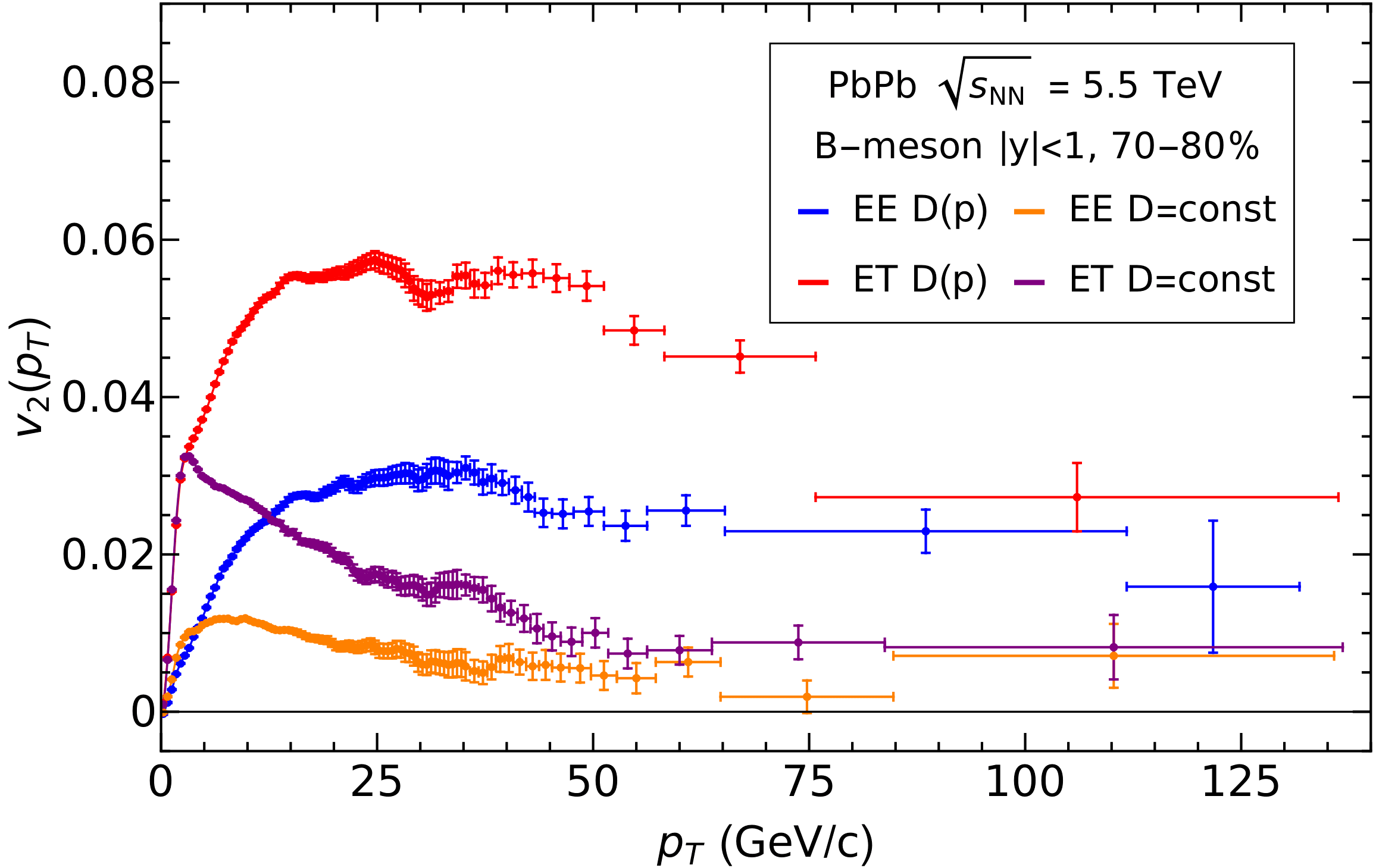} }}%
    \caption{\label{Fig7}B-meson $v_{2}(p_T)$ at $\sqrt{s_{NN}} =5.5$ TeV for the centrality class (\hyperref[Fig7a]{a}) 0-5\%, (\hyperref[Fig7b]{b}) 30-40\% and (\hyperref[Fig7c]{c}) 70-80\%.}%
\end{figure}

On the other hand, the momentum dependence of $R_{AA}(p_T)$ is a little bit harder to understand: the plots appear to show a stronger dependence on $p_T$ for the D=const scenario than the D(p) scenario.  The way to understand the unexpectedly stronger momentum dependence of the D=const scenario is to remember that $\mu$ in the D=const scenario is determined by the fluctuation-dissipation theorem, which is non-trivial for relativistic Brownian motion \cite{He:2013zua}.  Notice especially how the relativistic fluctuation-dissipation theorem leads to a drag coefficient that is inversely proportional to the energy of the heavy quark in the local fluid rest frame; see \eq{eq:dragconst}.  Thus higher-$p_T$ heavy quarks lose less energy in this scenario.  
Note further that the large mass of the $b$ quark implies that the $\gamma$ factor remains modest for experimentally accessible B meson momenta at the LHC, with a correspondingly modest increase in the size of momentum fluctuations for the D(p) scenario.

%%%%%%%%%%%%%%%%%%%%%%%%%%%%%%%%%%%%%%%%%%%%%%%%%%%%%%%%%%%%%%%%%%%%%%%%%%%%%%%%%%%%%%%%%%%%%%%%%%%%%%%%%%%%%%%%%%%%%%%%

\subsection{\label{v2} \texorpdfstring{$\mathbf{v_{2}(p_T)}$}{v2} at \texorpdfstring{$\mathbf{\sqrt{s_{NN}} =5.5}$}{v2} TeV}

We also present $v_{2}(p_T)$ results at $\surd s_{NN} =5.5$ TeV. In Fig.\ \ref{Fig6}, we show the centrality dependence of $v_{2}(p_T)$ for the EE parameter mapping with the diffusion coefficient that is dependent on momentum scenario, D(p). 
In order to increase readability, we show in Fig.\ \hyperref[Fig6]{6} (\hyperref[Fig6a]{a}) $v_2(p_T)$ for centrality classes 0-5\% up to 30-40\% and in Fig.\ \hyperref[Fig6]{6} (\hyperref[Fig6b]{b}) $v_2(p_T)$ for centrality classes 40-50\% up to 70-80\%. One can see that, as expected, $v_2(p_T)$ increases as one moves away from most central collisions as the initial geometrical asymmetry builds up.  The growth in $v_2(p_T)$ is non-monotonic, however, since in the more peripheral collisions the quarks have less medium and also a colder medium to propagate through; thus $v_2(p_T)$ decreases as a function of centrality from mid-central to peripheral collisions.  We show the $v_2(p_T)$ results from the other three setups in Appendix \ref{plots}.

%%%%%%%%%%%%%%%%%%%%%%%%%%%%%%%%%%%%%%%%%%%%%%%%%%%%%%%%%%%%%%%%%%%%%%%%%%%%%%%%%%%%%%%%%%%%%%%%%%%%%%%%%%%%%%%%%%%%%%
As was done for the $R_{AA}(p_T)$ results, we may also compare the $v_{2}(p_T)$ predictions for the four different setups---EE D(p), ET D(p), EE D=const, and ET D=const---for particular centrality classes.  We show this comparison in Fig.\ \ref{Fig7}.  In general, $R_{AA}$ and $v_2$ are anti-correlated: the greater the energy loss, the more quark momentum is reduced and the smaller the $R_{AA}$; at the same time, the greater the energy loss, the more sensitive quarks are to changes in geometry and thus the larger is $v_2$.  This anti-correlation is apparent in the ordering of $v_2(p_T)$ shown in Fig.\ \ref{Fig7}: the ET D=const setup has the largest $v_2(p_T)$ (smallest $R_{AA}(p_T)$) through the EE D(p) setup with the smallest $v_2(p_T)$ (largest $R_{AA}(p_T)$).  Further,
our $v_{2}(p_T)$ predictions at $\surd s_{NN} =5.5$ TeV for central collisions as shown in Fig.\ \hyperref[Fig7]{7} (\hyperref[Fig7a]{a}) are slightly higher compared to predictions made at $\surd s_{NN} = 2.76$ TeV shown in Fig.\ \ref{Fig3} for similar rapidities.  We can understand this slight increase in $v_2(p_T)$ with $\surd s_{NN}$ as due to the increased flow of the underlying medium.  

%%%%%%%%%%%%%%%%%%%%%%%%%%%%%%%%%%%%%%%%%%%%%%%%%%%%%%%%%%%%%%%%%%%%%%%%%%%%%%%%%%%%%%%%%%%%%%%%%%%%%%%%%%%%%%%%%%%%%%%

\subsection{Decoupling Energy Loss and Flow}

\begin{figure}[!htbp]
    \centering
    %\hspace*{-0.9cm}
    \subfigure[\label{subfig:Bv2IntOff}]{{\includegraphics[width=0.9\linewidth]{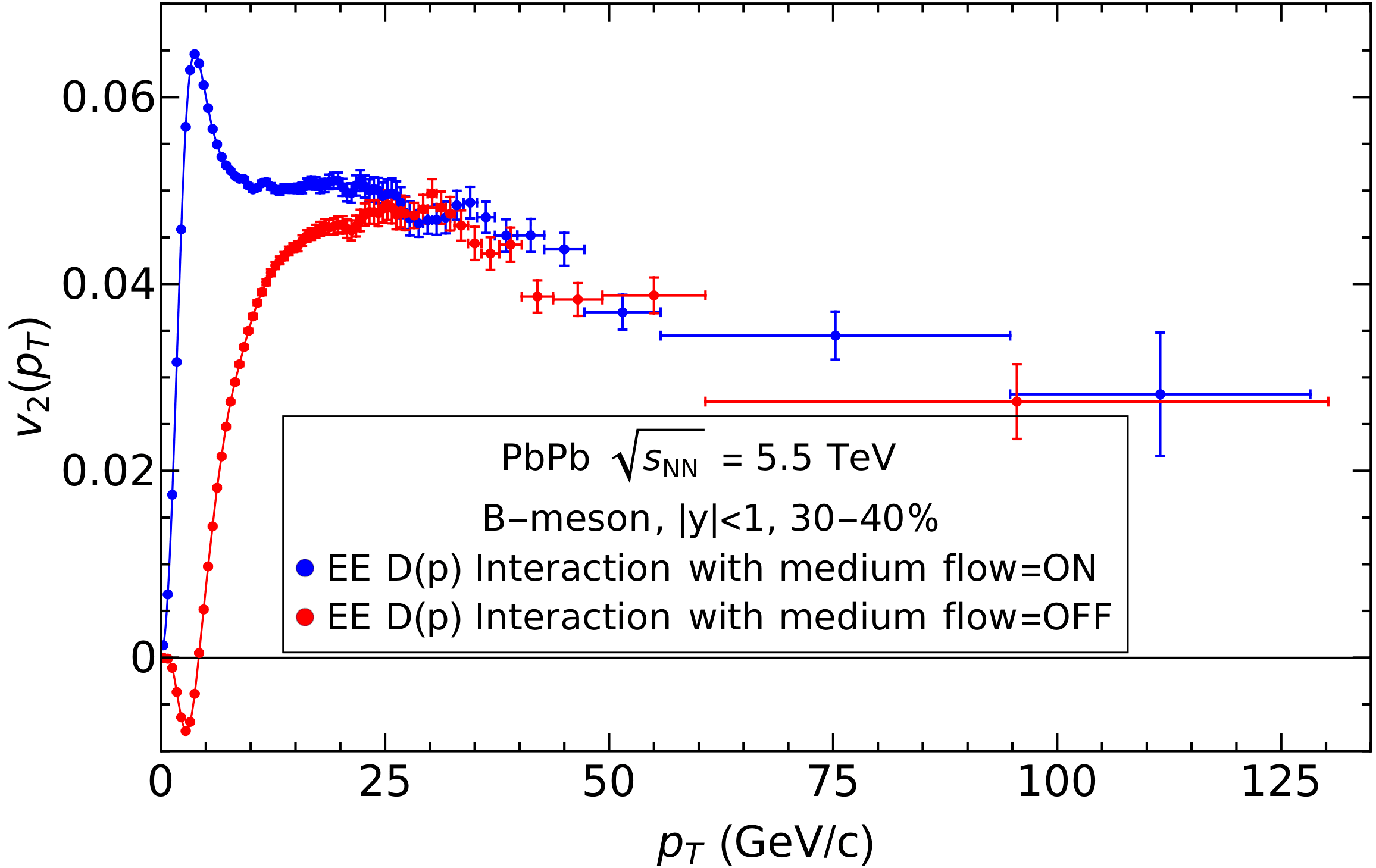} }}
    \caption{\label{fig:Bv2IntOff}B-meson $v_{2}(p_T)$ at $\sqrt{s_{NN}} =5.5$ TeV for the 30-40\% centrality class with the interaction between energy loss and flow on (blue) and off (red) for the EE D(p) scenario.}%
\end{figure}

To try to understand the interaction between the energy loss and the flow, we show a plot of $v_2(p_T)$ for 30-40\% centrality collisions in Fig.\ \ref{fig:Bv2IntOff} for the EE D(p) scenario for two cases: in blue we show the results for the unaltered model (cf.\ Fig.\ \hyperref[Fig6]{6}(\hyperref[Fig6a]{a})); in red, we decouple the energy loss from the flow by artificially turning off the boost to the local rest frame of the fluid for the energy loss step of the quark propagation.  Just to be clear: in the latter case the quark continues to propagate through the full VISHNU hydrodynamics background, which is expanding with time, and the quark continues to lose energy; however, the energy loss is computed in the lab frame, which is to say that during the energy loss part of the algorithm, the quark does not experience any effects of the velocity field of the hydrodynamics background---the quark doesn't ``feel'' the push or pull of the flowing medium.  

\section{\label{Conclusions}Conclusions and Outlook}

In this paper we presented predictions for the $R_{AA}(p_T)$ and $v_2(p_T)$ for B-mesons at $\surd s_{NN} =2.76$ TeV for central collisions and at $\surd s_{NN} =5.5$ TeV for central, semi-central and peripheral collisions from an energy loss model based on the AdS/CFT correspondence which assumes that the $b$ quarks are strongly coupled to a strongly coupled quark-gluon plasma.  We included predictions for two different ways to map the parameters in QCD to those in $\mathcal N = 4$ super Yang-Mills and also for two different scenarios for the momentum dependence of the diffusion coefficient.  These four different setups give a good idea of the systematic theoretical uncertainties in current, state-of-the-art phenomenological energy loss models based on AdS/CFT and provide predictions for comparison to future measurements from the LHC.  

Our results for B meson $R_{AA}(p_T)$ and $v_2(p_T)$ at $\surd s_{NN} = 2.76$ TeV are consistent with our previous predictions \cite{Horowitz:2015dta}, and are in qualitative agreement with CMS data \cite{Khachatryan:2016ypw,Sirunyan:2017oug}.

Our results generally behave as expected as a function of centrality: higher centrality leads to less suppression and a larger $R_{AA}(p_T)$ due to the smaller, cooler, shorter-lived medium.  The anisotropy $v_2(p_T)$ increases as a function of centrality from most-central collisions to mid-central collisions, as the geometrical anisotropy and medium momentum flow build up.  Beyond mid-central collisions the nuclear overlap has less initial geometrical anisotropy and the smaller, cooler, shorter-lived QGP implies less energy loss and therefore less sensitivity to the medium geometry; hence $v_2(p_T)$ decreases from mid-central to peripheral collisions.  

The nuclear modification factor also generally behaves as expected as a function of momentum.  For $p_T\lesssim 5$ GeV/c, $R_{AA}$ grows larger than 1 due to conservation of $b$ quark number.  The $R_{AA}(p_T)$ then rapidly decreases as a function of $p_T$ due to energy loss, followed by a gradual increase as a function of momentum.  For the D(p) scenario, $R_{AA}(p_T)$ increases with $p_T$ due to the increasing size of the momentum fluctuations due to the dependence of the momentum diffusion coefficient predicted in \cite{Gubser:2006nz}.  For the D=const scenario, $R_{AA}(p_T)$ increases with $p_T$ since the relativistic fluctuation-dissipation theorem requires that the drag $\mu\sim1/E$ for a momentum independent diffusion coefficient \cite{He:2013zua}; therefore higher momentum $b$ quarks have a smaller drag coefficient than lower momentum $b$ quarks.

The azimuthal anisotropy of the unaltered energy loss model can also be understood qualitatively.  At low momentum, $v_2\rightarrow0$ as $p_T\rightarrow0$: as the quark is (essentially) stopped in a short distance, the quark is insensitive to the geometry; additionally, momentum fluctuations further wash out any sensitivity to the geometry.  As the momentum increases, the quark becomes sensitive to the geometry, both in terms of the quantity of medium through which the quark propagates as well as the medium's flow.  For higher momenta, the $v_2(p_T)$ decreases with momentum as the energy loss decreases with momentum; generally speaking, $v_2$ and $R_{AA}$ are anti-correlated.  

We saw a huge qualitative difference between the $v_2$ predictions from the unaltered energy loss model and from the energy loss model in which the energy loss was decoupled from the flow.  This huge difference was initially surprising, but can be understood with some thought.  With the flow interaction turned off, for quarks moving slower than the medium speed of $\sim \mathrm{c}/2$, the medium in the $\phi\sim0$ direction actually gets larger with time than the medium in the $\phi\sim\pi/2$ direction.  As before, as $p_T\rightarrow0$ one must have $v_2\rightarrow0$ as the quark simply stops.  Hence $v_2(p_T)$ \emph{decreases} from 0 as $p_T$ increases until $p_T\sim M_Q$..  As $p_T$ increases beyond $\sim M_Q$, the quark moves ever faster than the medium expansion and $v_2(p_T)$ increases with $p_T$.  The increase of $v_2$ with $p_T$ continues until it plateaus at its maximum value between $0.045-0.050$; then $v_2(p_T)$ decreases with $p_T$ as the momentum fluctuations become larger and $R_{AA}$ increases with $p_T$ starting from $p_T\sim25-30$ GeV/c.  

For the usual case in which the quark energy loss is coupled to the flow, we see that $v_2(p_T)$ increases with $p_T$ until again $p_T\sim M_Q$.  In this case, when the quark is propagating along $\phi\sim0$ for $p_T\lesssim M_Q$ the medium is pushing the quark along.  When the velocity of the quark is the same as that of the plasma, there is on average no energy loss at all; hence $v_2$ reaches a maximum when $p_T\sim M_Q$.  As $p_T$ continues to increase the quark begins to lose energy in all directions of motion, and thus $v_2$ decreases as a function of $p_T$.  The width of this decrease should be, and is, roughly the width of the $p_T$ required to reach the maximum $v_2$.

What \emph{is} very surprising is just how far out in $p_T$ the difference between the two cases continues; the two different $v_2$'s only converge at $p_T\sim25-30$ GeV/c, when the momentum fluctuations lead to $v_2(p_T)$ generally decreasing with $p_T$.  There's no obvious natural momentum scale in the problem to suggest that the effects of flow on $v_2$ generated by energy loss should persist to momenta this large.  What's so tantalizing about this result is that it may point to a resolution to the 20-year-long $R_{AA}$, $v_2$ puzzle at intermediate-$p_T$ for light hadrons \cite{Shuryak:2001me}.  Rather than requiring exotic, difficult to control, non-perturbative physics such as magnetic monopoles \cite{Xu:2014tda} or changes to hadronization \cite{He:2011qa}, the surprisingly large $v_2$ of light hadrons at $p_T\sim5-15$ GeV/c \cite{Horowitz:2012cf} might be due simply to as-yet uncomputed and unquantified interaction between the energy loss of light partons and the medium flow.  

The qualitative agreement between our predictions and data \cite{Horowitz:2015dta,Hambrock:2017sno} and the large systematic theoretical uncertainties in the energy loss predictions due to our current limits of understanding of energy loss at strong coupling argue for continued research into the phenomenology of the AdS/CFT correspondence.

In this paper, we've performed phenomenological calculations for B mesons that can be compared to data. One can also perform these calculations for D-mesons as well as other collision systems such as $Xe+Xe$ \cite{Acharya:2018eaq} and this is left for future work.  

Our energy loss model only considers heavy quark propagation beyond the thermalization time given by VISHNU; incorporating pre-thermalization energy loss effects could provide insight on the collision medium prior to the applicability of hydrodynamics.  Since the initial production of high-$p_T$ particles is described by pQCD, using a pQCD energy loss model before thermalization followed by a strong coupling treatment post thermalization may be a reasonable approach.  One could also investigate whether AdS/CFT energy loss calculations can be applied to low energy heavy-ion collisions, although this will likely require one to account for the non-zero baryon chemical potential and low temperature effects on the drag and diffusion terms.

\begin{acknowledgments}
The authors wish to acknowledge Valumax Projects (Pty) Ltd, SA-CERN, The South African National Research Foundation (NRF), Professor Amanda Weltman and the University of Cape Town through Form 10A funding, the Vice Chancellor's Research scholarship as well as the SA College Croll scholarship for their generous financial contributions towards this work.
\end{acknowledgments}

\appendix

\section{Geometric quantities in the Glauber optical limit}

The binary nucleon-nucleon collision density in the transverse plane is given by \cite{Miller:2007ri},
\label{eq:collden}
\begin{eqnarray}
n_{BC} (x,y;b) =&& AB \sigma_{inel}^{NN} T_A \left( x-\frac{b}{2}, y \right) \nonumber \\
&&\times T_B \left( x+\frac{b}{2}, y \right) 
\end{eqnarray}

\noindent
where $A$ and $B$ are the number of nucleons in nucleus A and B respectively, $b$ is the impact parameter and the definition of $T_{A/B}$ is given in \cite{Miller:2007ri}. 

The geometric cross section distribution with respect to the impact parameter $d\sigma/db$ is given by
\label{eq:dsigma}
\begin{eqnarray}
\frac{d\sigma}{db} (b)  &=& 2 \pi b (1-[1-T_{AB}(b)\sigma_{inel}^{NN}]^{AB}),
\end{eqnarray}
and centrality classes are defined theoretically by taking slices of this distribution. Table \ref{tab:centrality} shows the centrality classes we've employed for $Pb+Pb$ at $\surd s_{NN}=5.5$ TeV. The impact parameter values that we've provided are consistent within 2\% of the results in Ref.\ \cite{Loizides:2017ack}.

\begin{table}[!htbp]
\caption{\label{tab:centrality} Centrality classes for $Pb+Pb$ at $\sqrt{s_{NN}}=5.5$ TeV}
\begin{ruledtabular}
\begin{tabular}{cccc}
\textrm{Centrality}&
\textrm{$b_{min}$ (fm)}&
\textrm{$b_{max}$ (fm)}&
\textrm{$\langle b \rangle$ (fm)}\\
\colrule
 0-5\%             &   0            &   3.55    & 2.37\\
    5-10\%             &   3.55            &   5.02    & 3.35\\
    10-20\%             &   5.02            &   7.01    & 6.12\\
    20-30\%             &   7.01            &   8.70    & 7.92\\
    30-40\%             &   8.70            &   10.04    & 9.38\\
    40-50\%             &   10.04            &   11.23    & 10.64\\
    50-60\%             &   11.23            &   12.30    & 11.77\\
    60-70\%             &   12.30            &   13.28    & 12.80\\
    70-80\%             &   13.28            &   14.20    & 13.75\\
    80-90\%             &   14.20            &   15.12    & 14.65\\
    90-100\%             &   15.12            &   21.69    & 15.95\\
\end{tabular}
\end{ruledtabular}
\end{table}

\section{\label{plots}Plots of additional results at \texorpdfstring{$\mathbf{\sqrt{s_{NN}} =5.5}$}{plots} TeV}

\begin{figure*}[!htbp]
    \centering
    %\hspace*{-0.9cm}
    \subfigure[\label{Fig8a}]{{\includegraphics[width=0.47\linewidth]{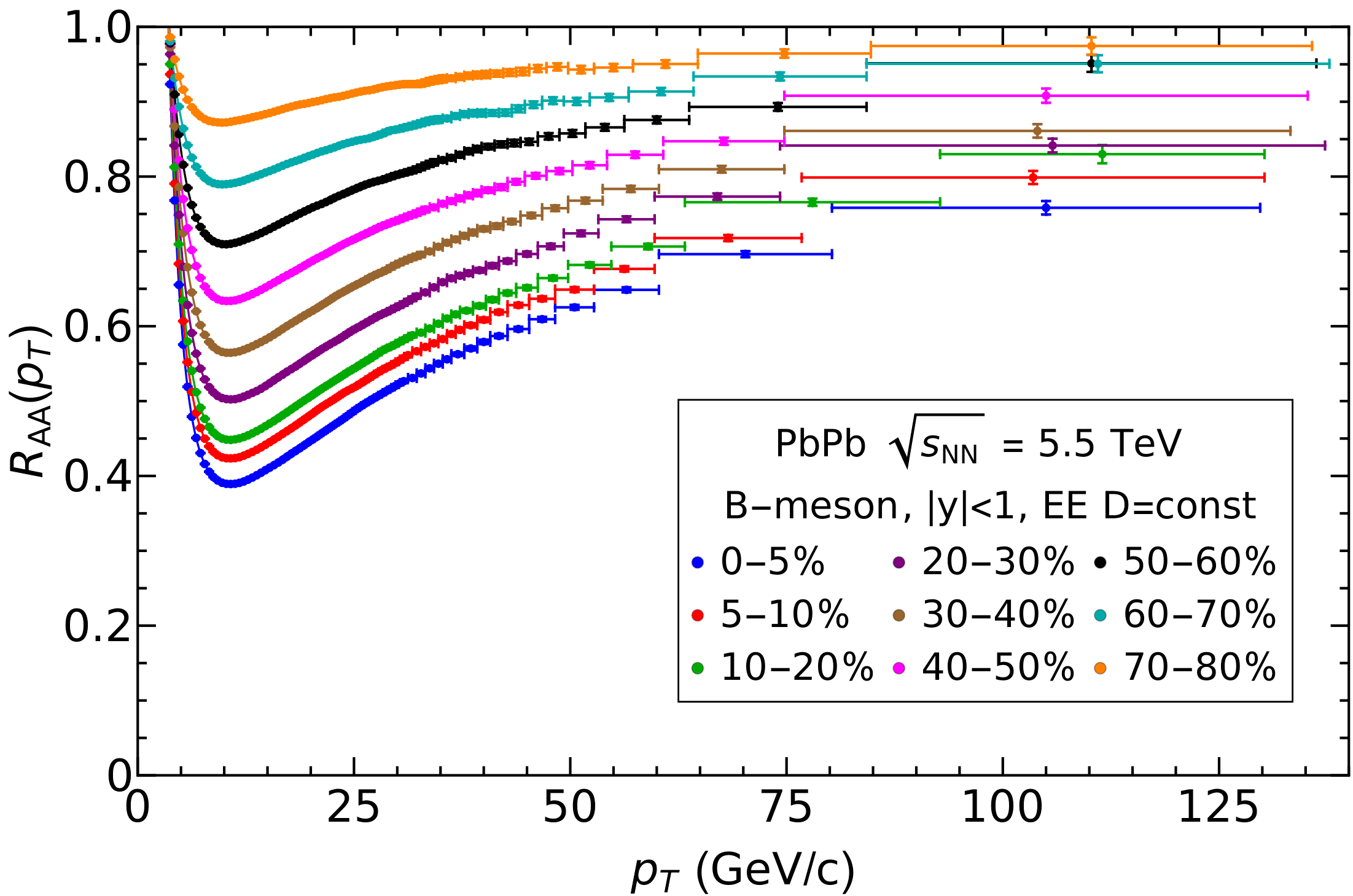} }}%
    \hspace{1.5em}
    \subfigure[\label{Fig8b}]{{\includegraphics[width=0.47\linewidth]{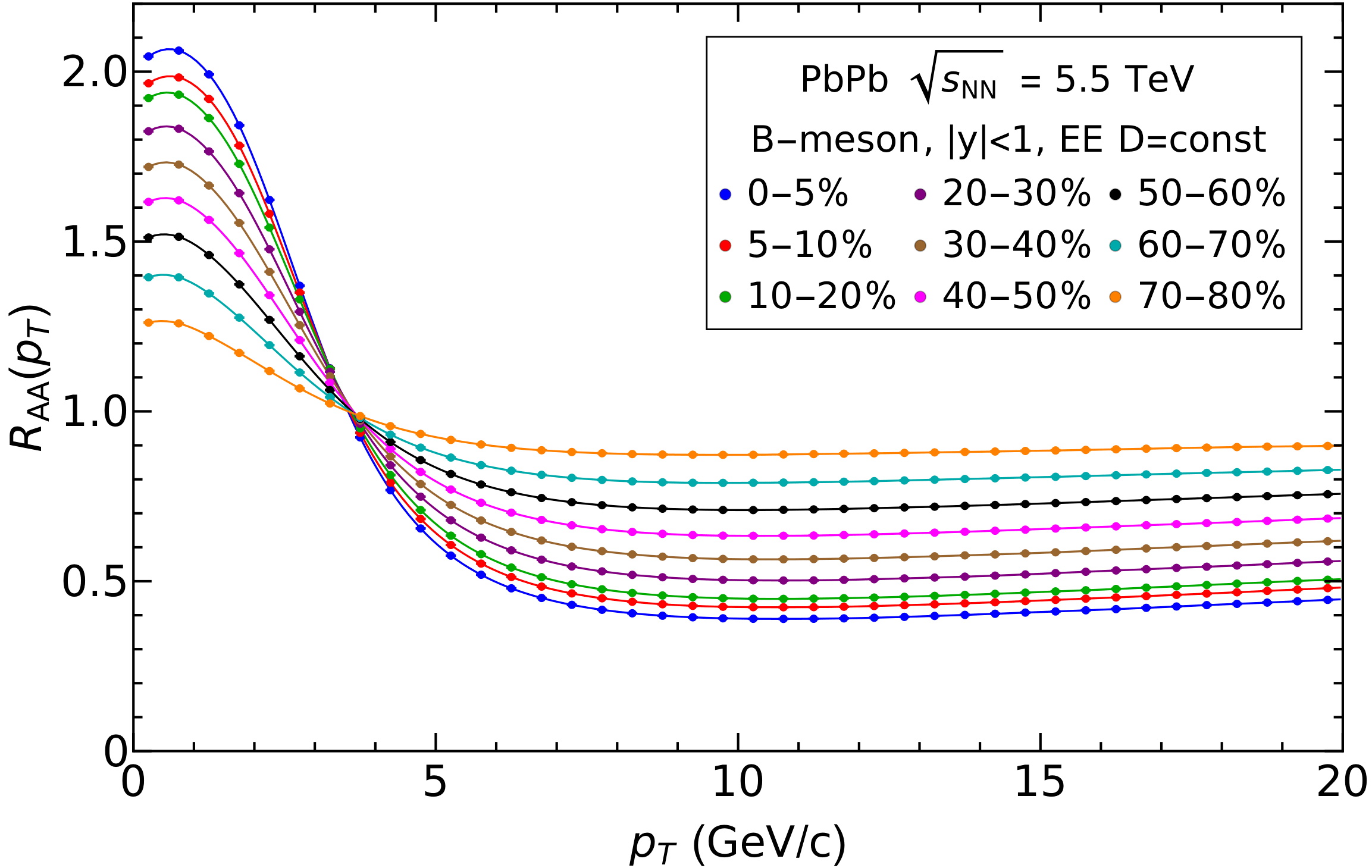} }}%
    \caption{\label{Fig8}(\hyperref[Fig8a]{a}) EE, D=const B-meson $R_{AA}(p_T)$ at $\sqrt{s_{NN}} =5.5$ TeV for centrality classes 0-5\% up to 70-80\%. (\hyperref[Fig8b]{b}) Expanded view of the transverse momentum region, $0<p_T \leq 20$ GeV/c of (a), including the region $R_{AA}(p_T)>1$.}%
\end{figure*}

For completeness and future reference, we provide the remaining sets of predictions of B meson $R_{AA}(p_T)$ and $v_2(p_T)$ for $Pb+Pb$ collisions at $\surd s_{NN}=5.5$ TeV from all setups of our energy loss model in Figs.\ \hyperref[Fig8]{8} - \hyperref[Fig13]{13}.

\onecolumngrid
% \begin{widetext}

\begin{figure*}[!htbp]
    \centering
    %\hspace*{-0.9cm}
    \subfigure[\label{Fig9a}]{{\includegraphics[width=0.47\linewidth]{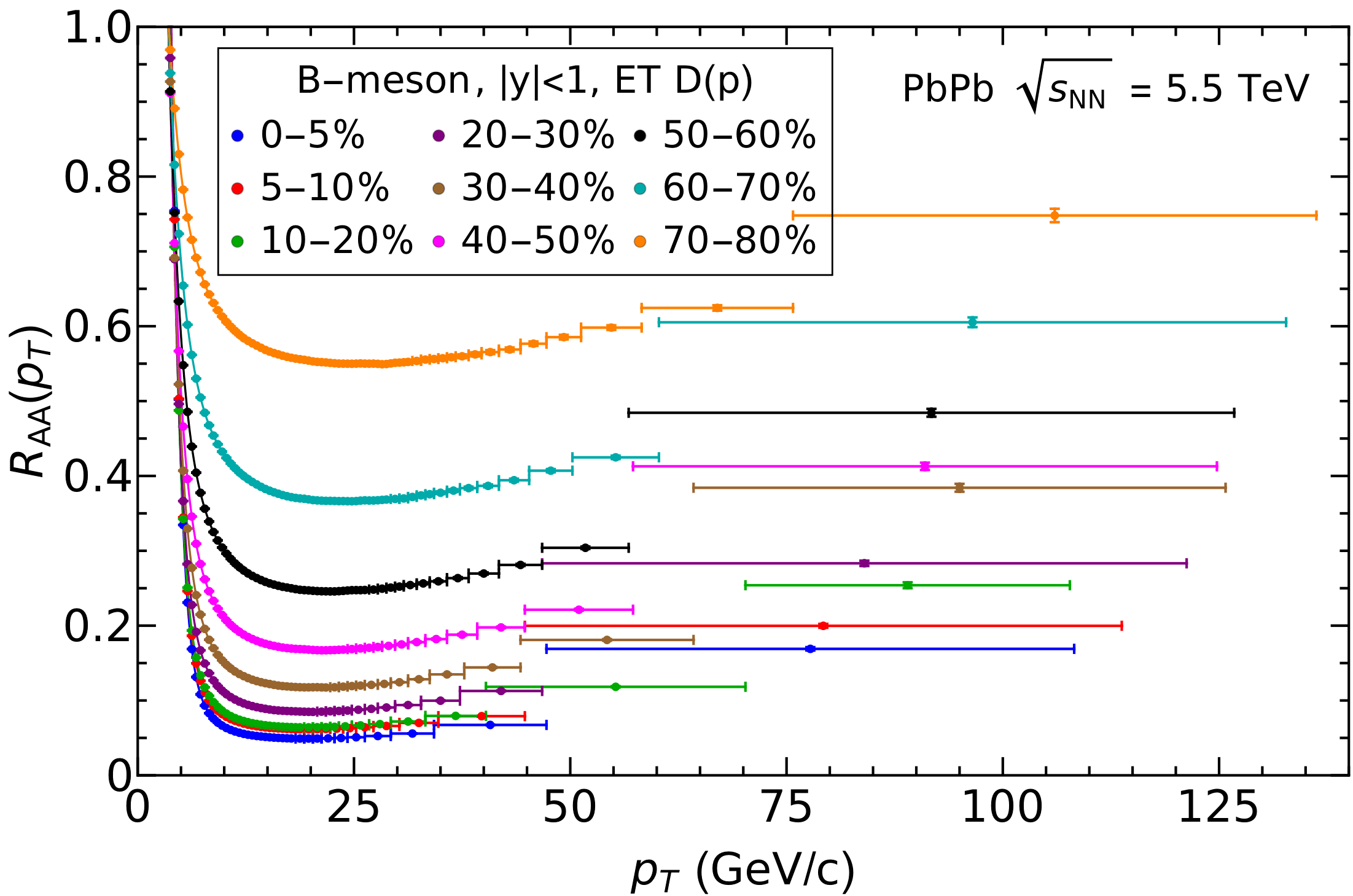} }}%
    \hspace{1.5em}
    \subfigure[\label{Fig9b}]{{\includegraphics[width=0.47\linewidth]{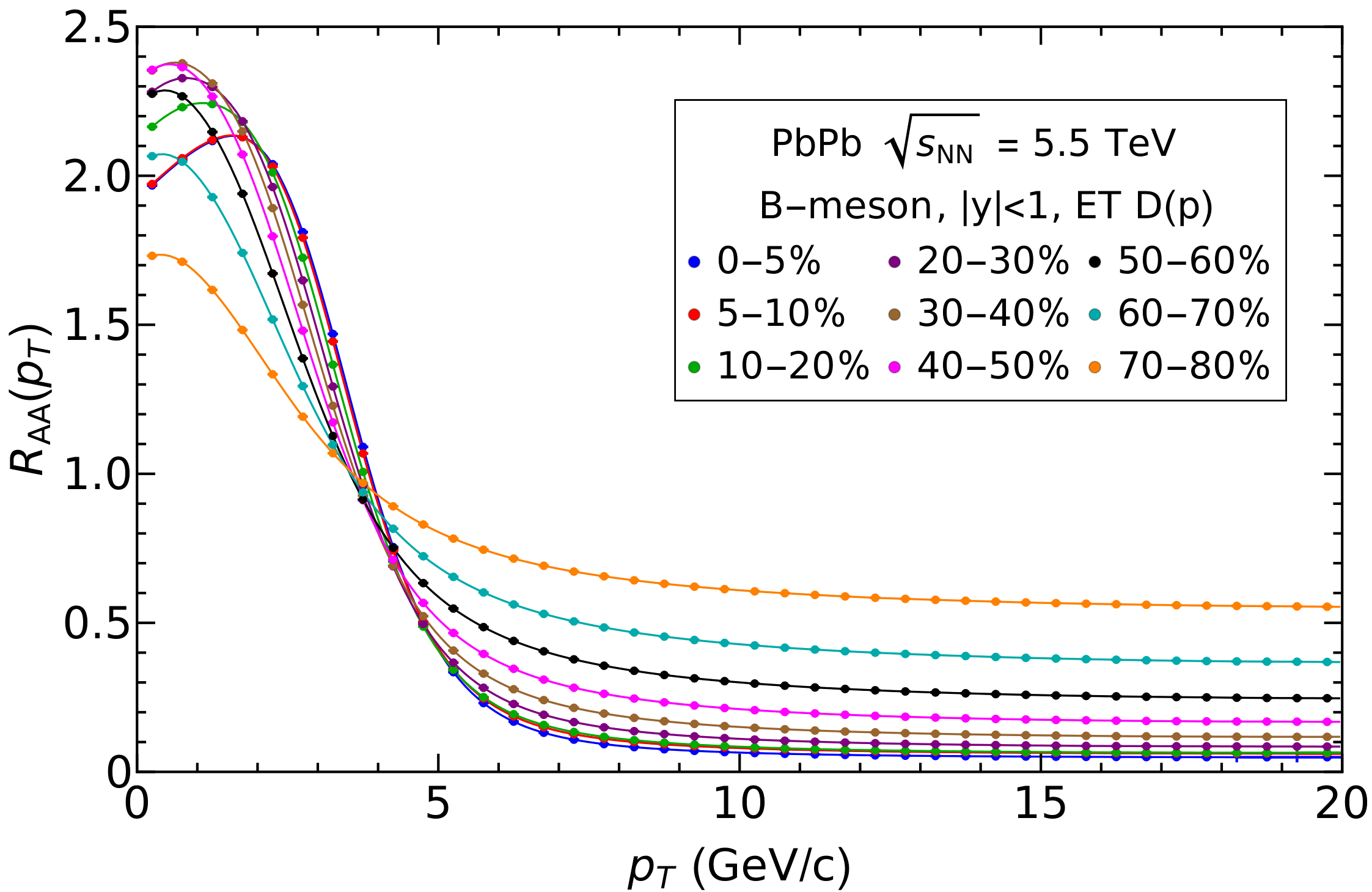} }}%
    \caption{\label{Fig9}(\hyperref[Fig9a]{a}) ET, D(p) B-meson $R_{AA}(p_T)$ at $\sqrt{s_{NN}} =5.5$ TeV for centrality classes 0-5\% up to 70-80\%. (\hyperref[Fig9b]{b}) Expanded view of the transverse momentum region, $0<p_T \leq 20$ GeV/c of (a), including the region $R_{AA}(p_T)>1$.}%
\end{figure*}

\begin{figure*}[!htbp]
    \centering
    %\hspace*{-0.9cm}
    \subfigure[\label{Fig10a}]{{\includegraphics[width=0.47\linewidth]{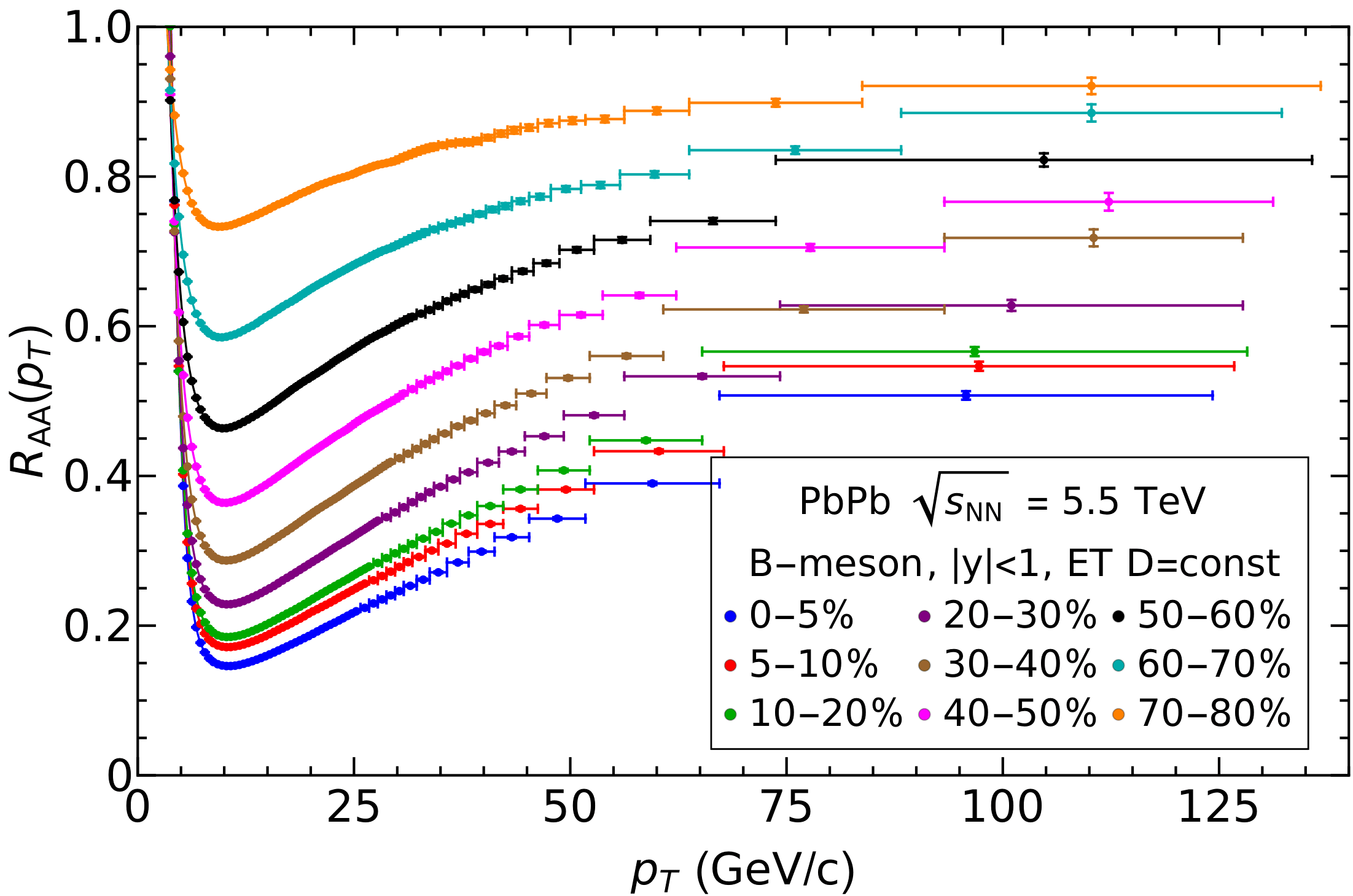} }}%
    \hspace{1.5em}
    \subfigure[\label{Fig10b}]{{\includegraphics[width=0.47\linewidth]{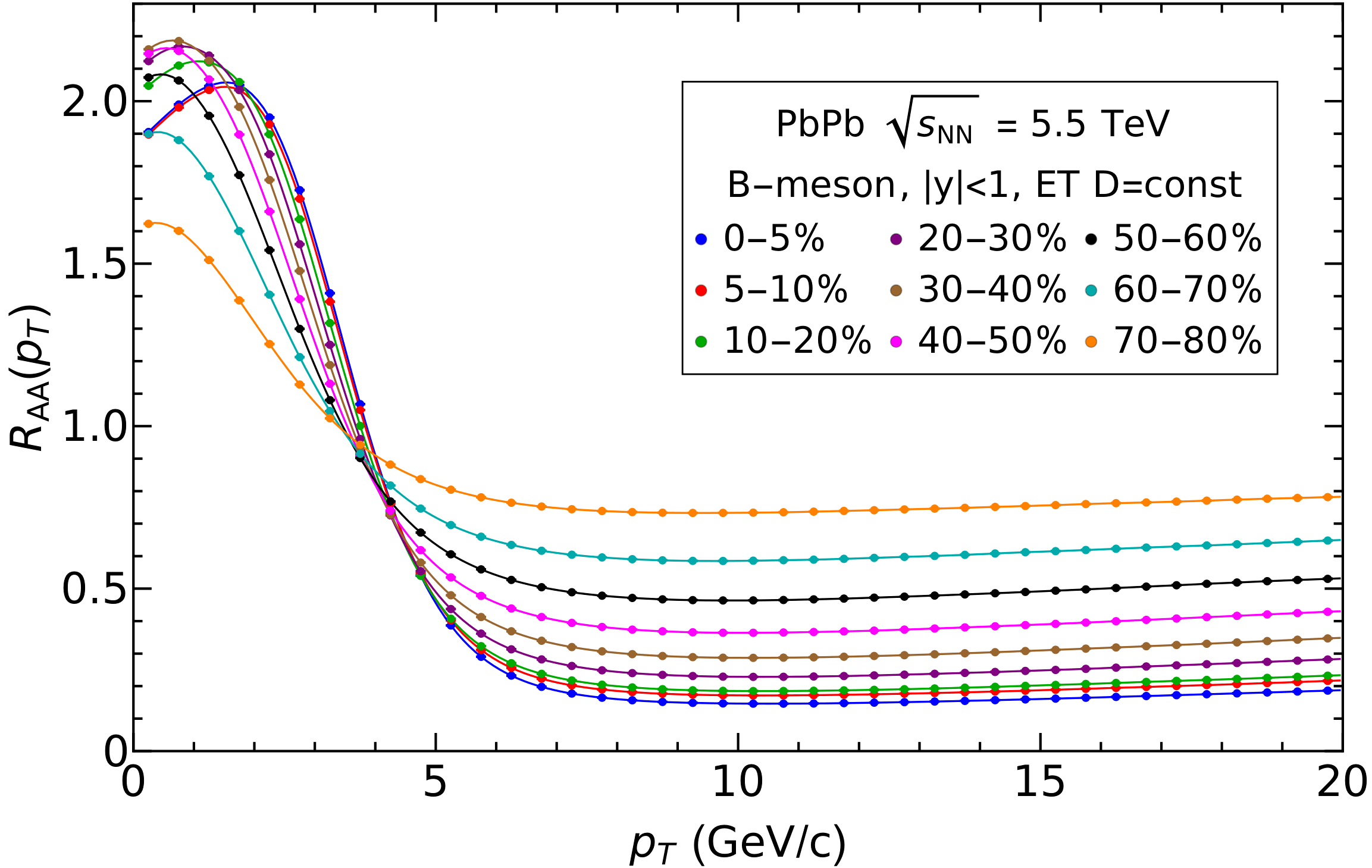} }}%
    \caption{\label{Fig10}(\hyperref[Fig10a]{a}) ET, D=const B-meson $R_{AA}(p_T)$ at $\sqrt{s_{NN}} =5.5$ TeV for centrality classes 0-5\% up to 70-80\%. (\hyperref[Fig10b]{b}) Expanded view of the transverse momentum region, $0<p_T \leq 20$ GeV/c of (a), including the region $R_{AA}(p_T)>1$.}%
\end{figure*}

%%%%%%%%%%%%%%%%%%%%%%%%%%%%%%%%%%%%%%%%%%%%%%%%%%%%%%%%%%%%%%%%%%%%%%%%%%%%%%%%%%%%%%%%%%%%%%%%%%%%%%%%%%%%%%%%%%%%%%

%\subsection{\label{v2_apx}$\mathbf{v_{2}(p_T)}$}

\begin{figure*}[!htbp]
    \centering
    %\hspace*{-0.9cm}
    \subfigure[\label{Fig11a}]{{\includegraphics[width=0.47\linewidth]{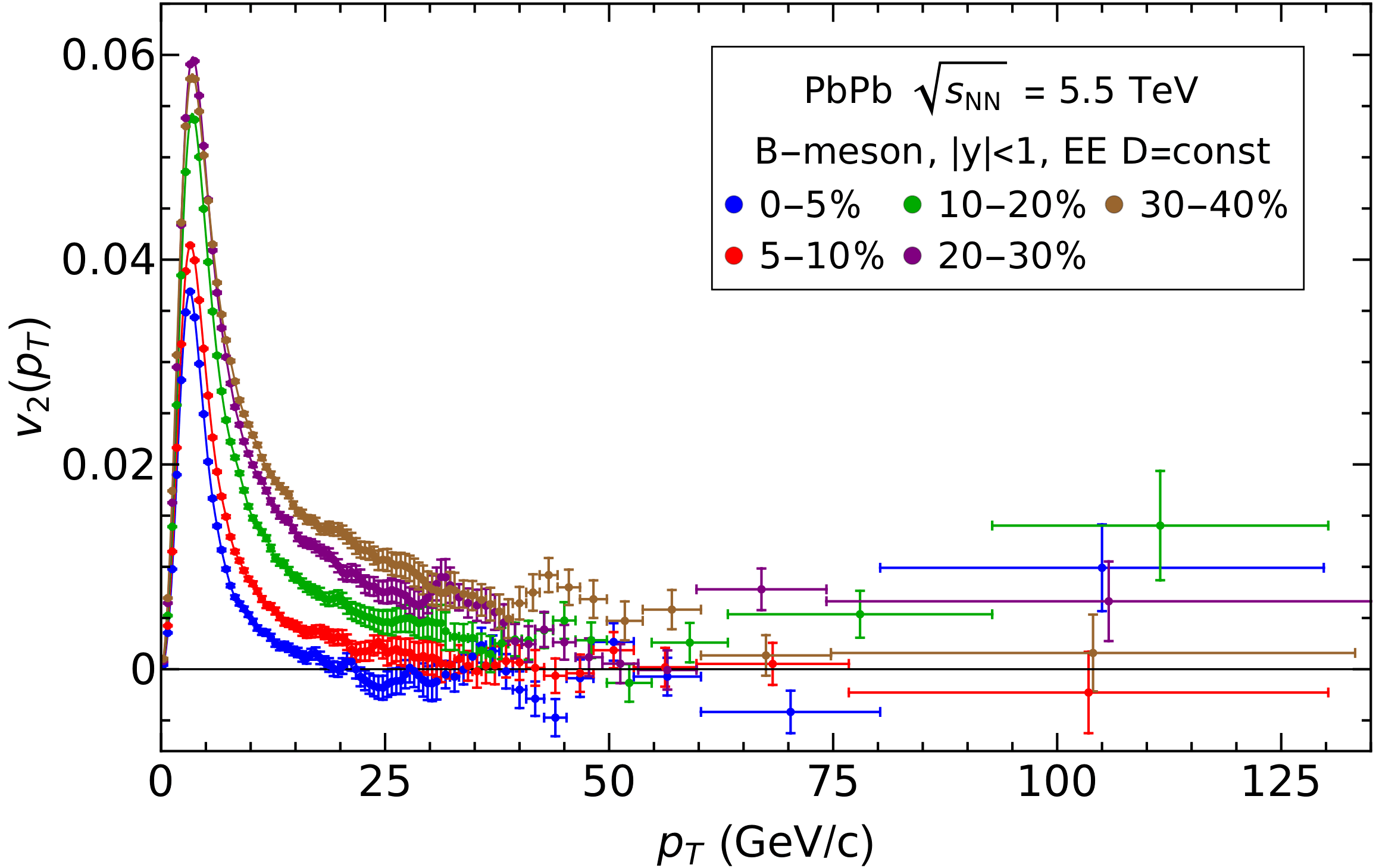} }}%
    \hspace{1.5em}
    \subfigure[\label{Fig11b}]{{\includegraphics[width=0.47\linewidth]{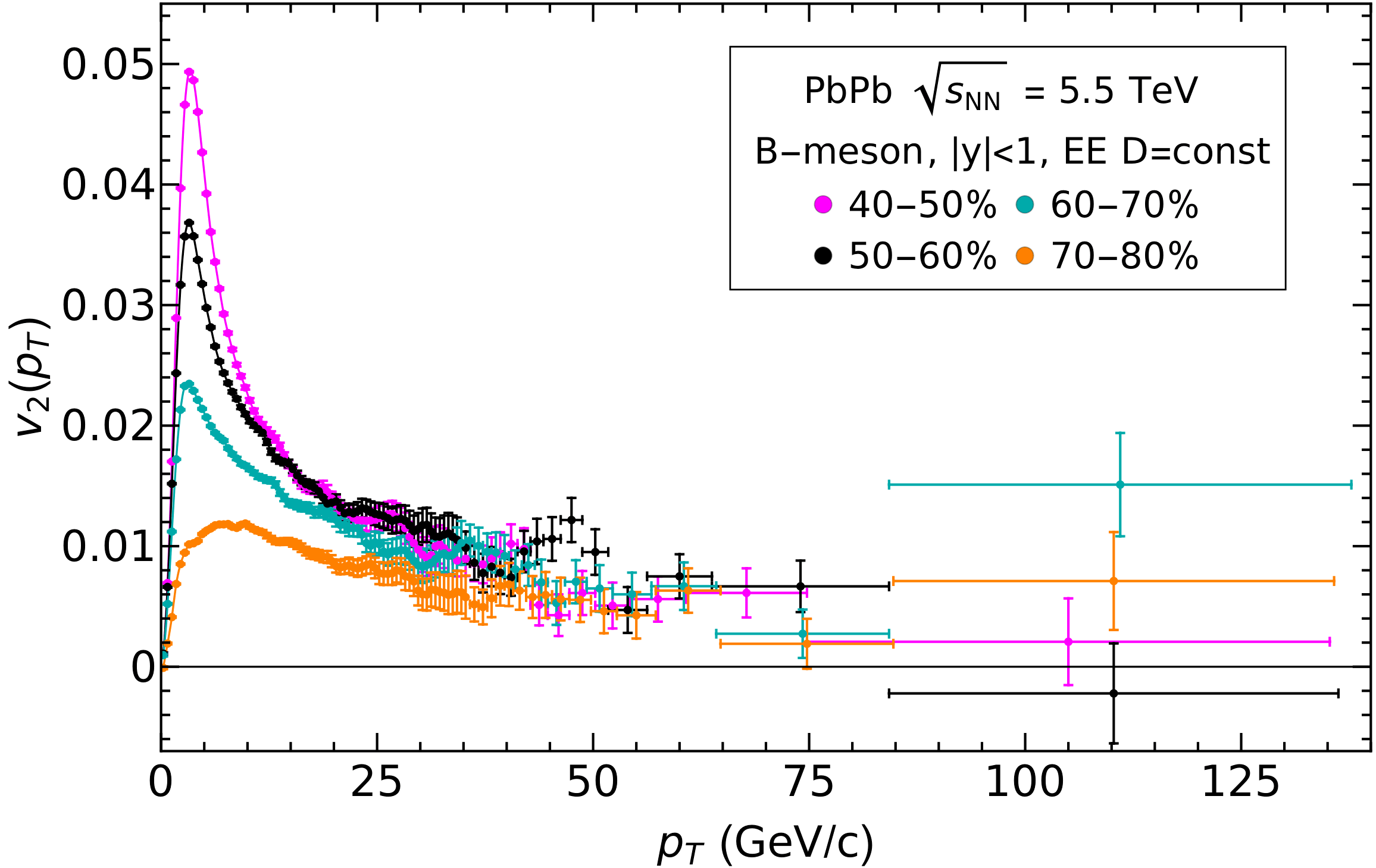} }}%
    \caption{\label{Fig11}EE, D=const B-meson $v_{2}(p_T)$ at $\sqrt{s_{NN}} =5.5$ TeV for the centrality classes (\hyperref[Fig11a]{a}) 0-5\% up to 30-40\% and (\hyperref[Fig11b]{b}) 40-50\% up to 70-80\%.}%
\end{figure*}

\begin{figure*}[!htbp]
    \centering
    %\hspace*{-0.9cm}
    \subfigure[\label{Fig12a}]{{\includegraphics[width=0.47\linewidth]{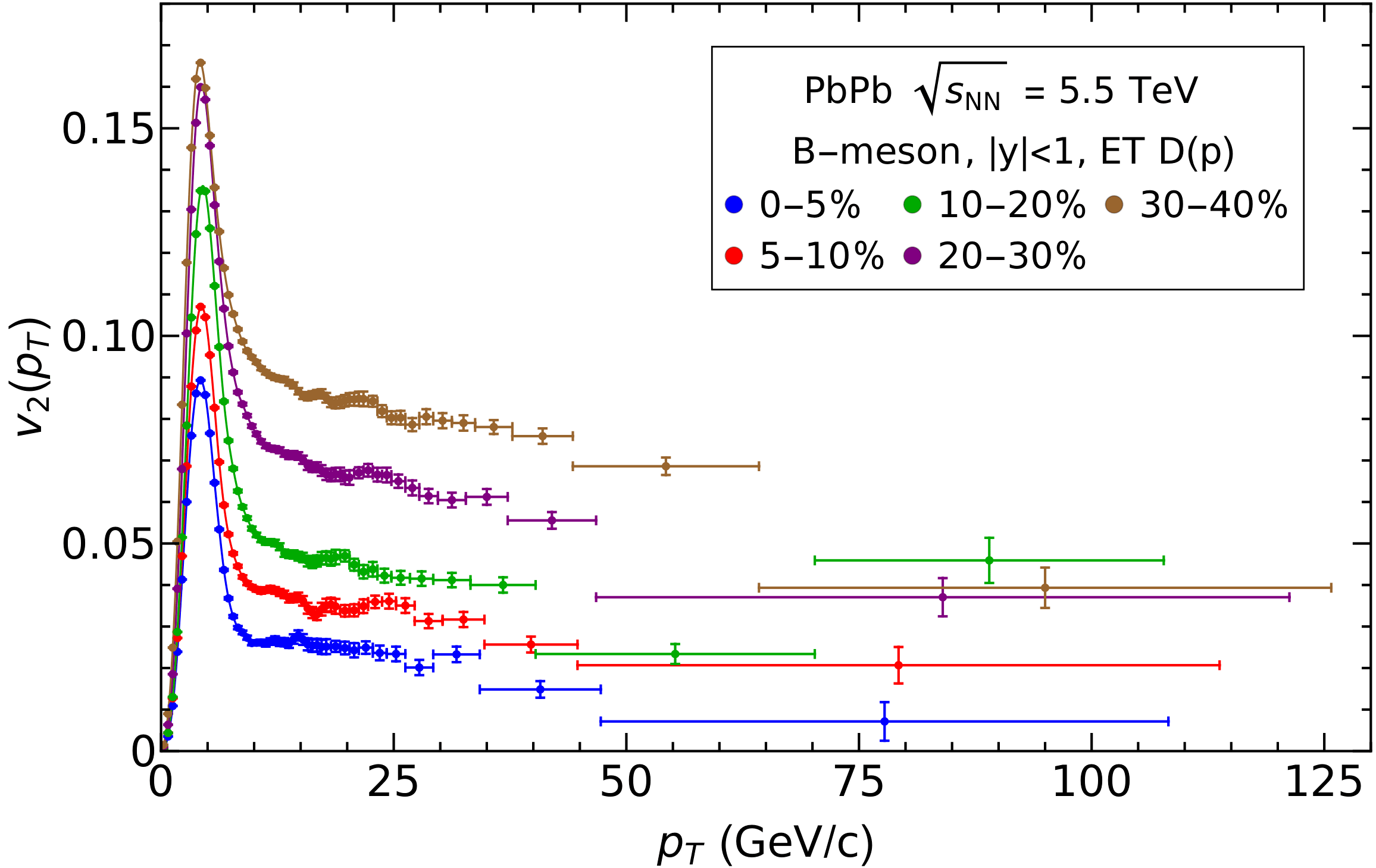} }}%
    \hspace{1.5em}
    \subfigure[\label{Fig12b}]{{\includegraphics[width=0.47\linewidth]{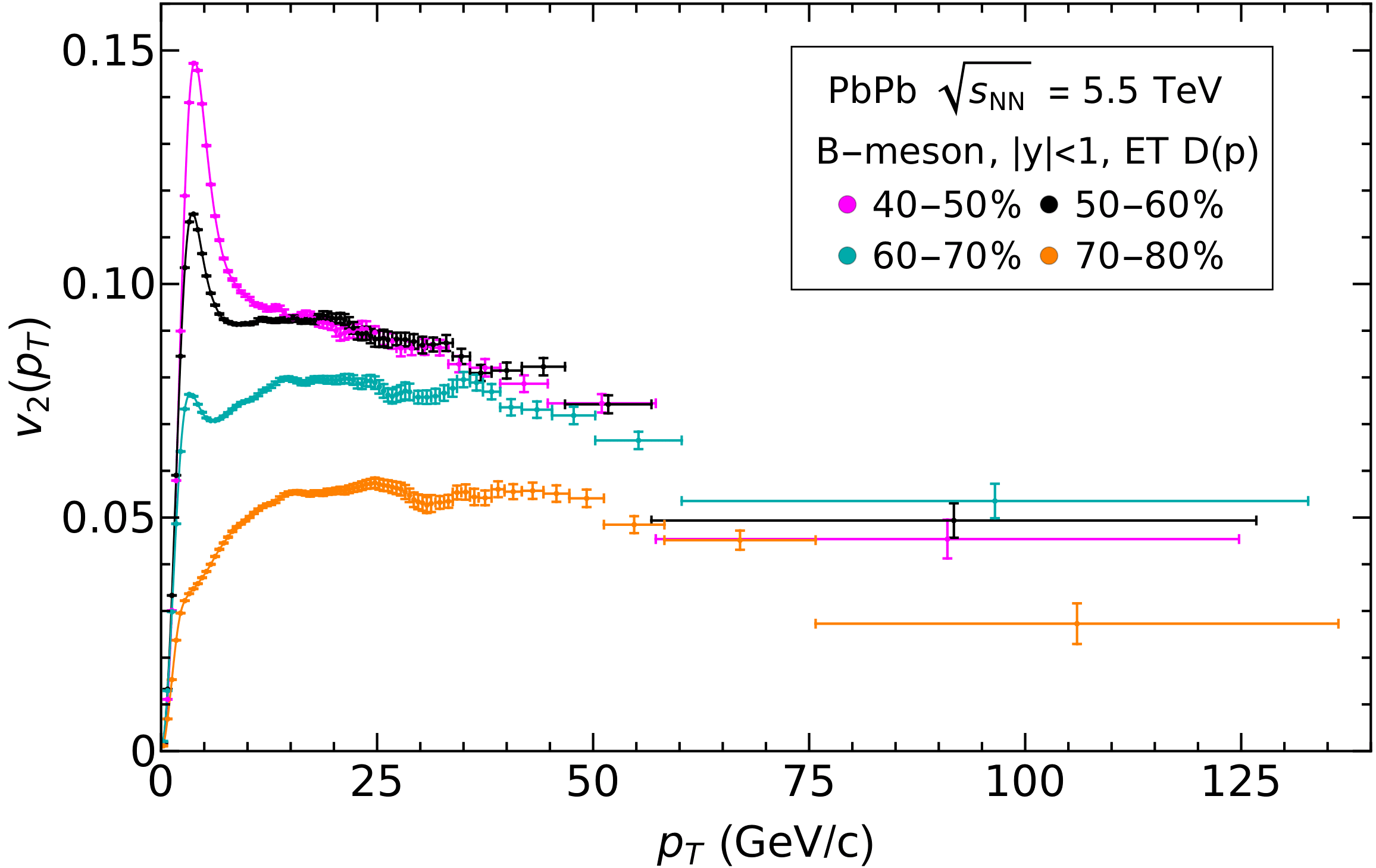} }}%
    \caption{\label{Fig12}ET, D(p) B-meson $v_{2}(p_T)$ at $\sqrt{s_{NN}} =5.5$ TeV for the centrality classes (\hyperref[Fig12a]{a}) 0-5\% up to 30-40\% and (\hyperref[Fig12b]{b}) 40-50\% up to 70-80\%.}%
\end{figure*}

\begin{figure*}[!htbp]
    \centering
    %\hspace*{-0.9cm}
    \subfigure[\label{Fig13a}]{{\includegraphics[width=0.47\linewidth]{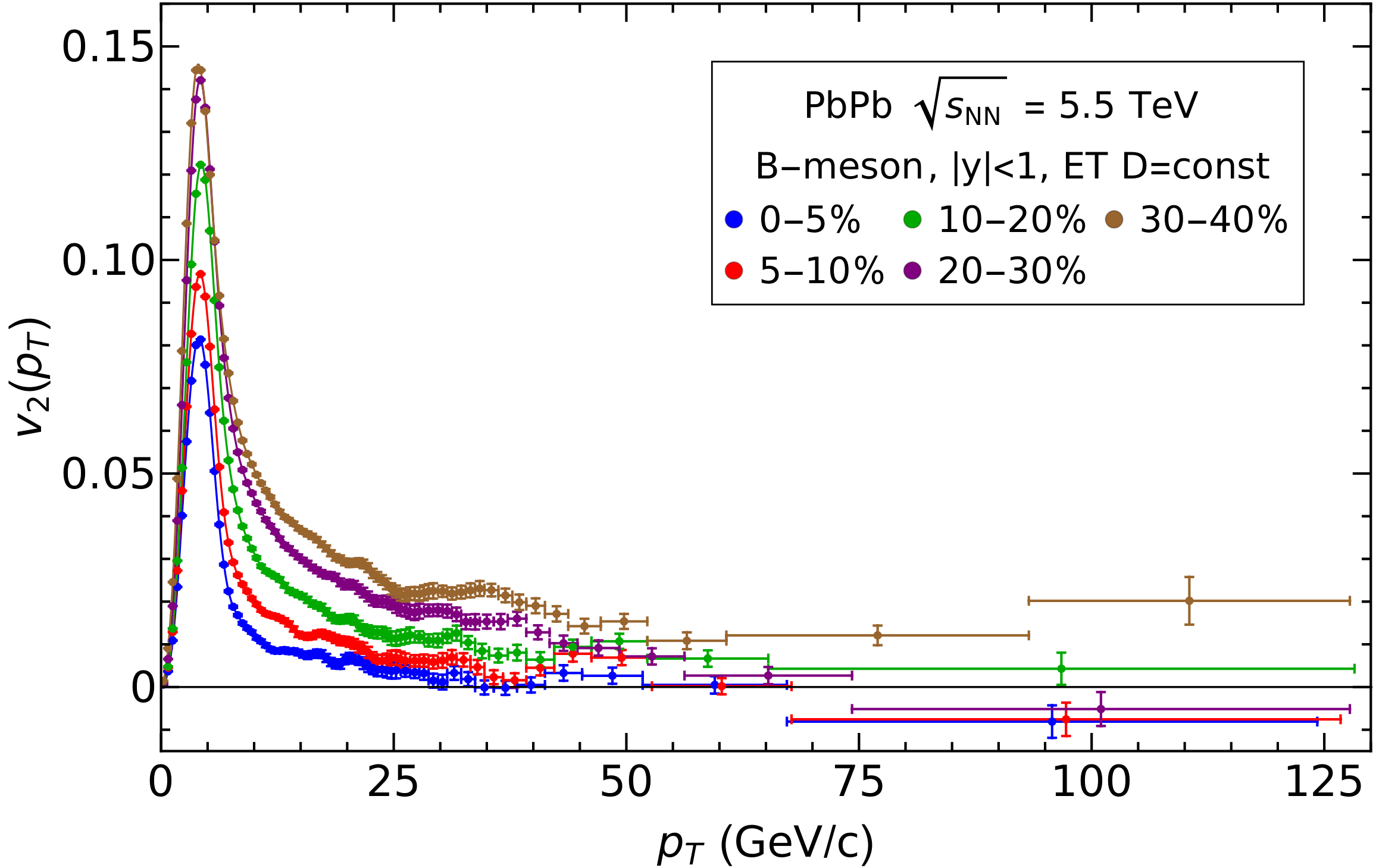} }}%
    \hspace{1.5em}
    \subfigure[\label{Fig13b}]{{\includegraphics[width=0.47\linewidth]{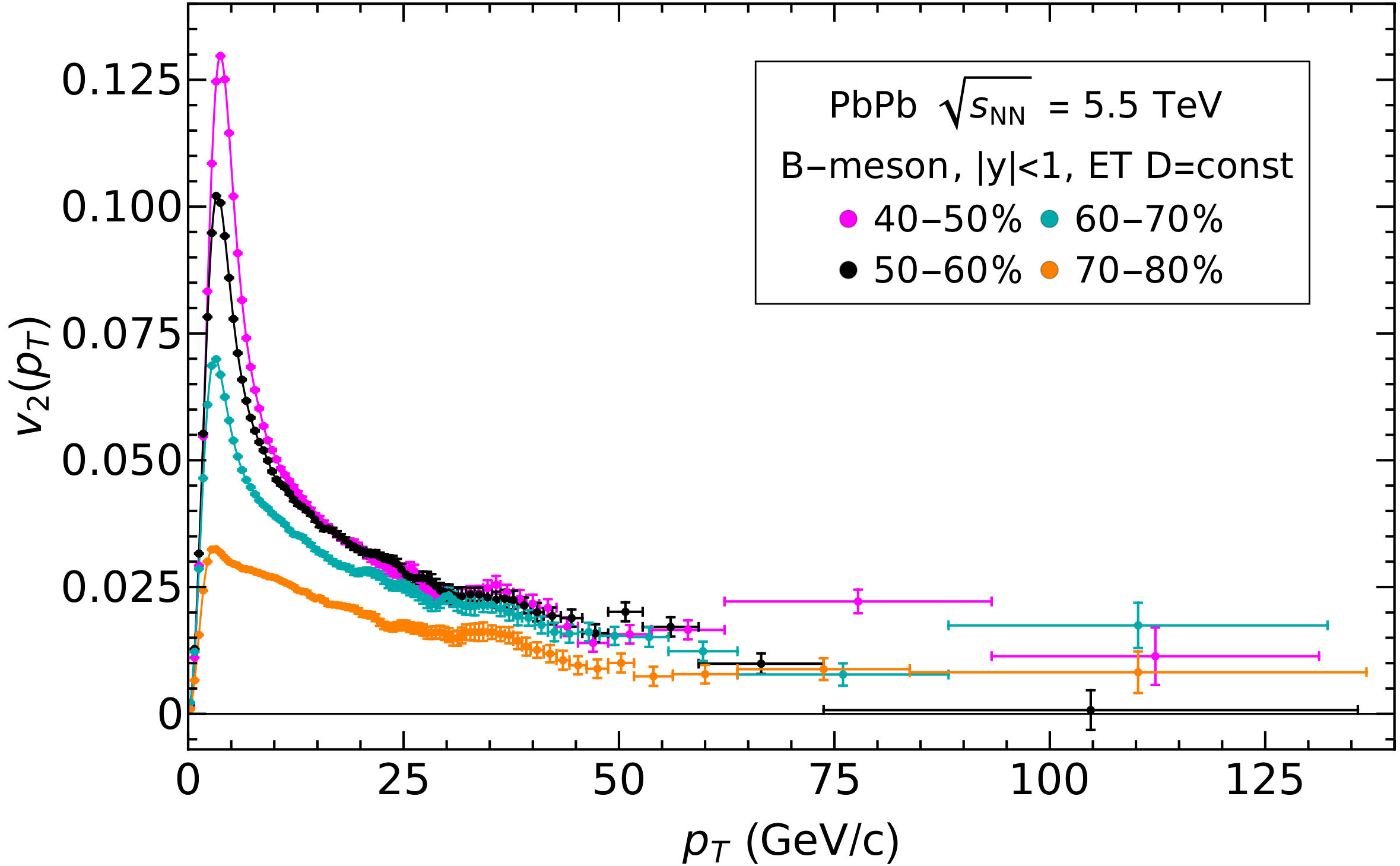} }}%
    \caption{\label{Fig13}ET, D=const B-meson $v_{2}(p_T)$ at $\sqrt{s_{NN}} =5.5$ TeV for the centrality classes (\hyperref[Fig13a]{a}) 0-5\% up to 30-40\% and (\hyperref[Fig13b]{b}) 40-50\% up to 70-80\%.}%
\end{figure*}

% \end{widetext}

\FloatBarrier %Block figures from going into the appendix

\twocolumngrid

% \bibliography{apssamp}% Produces the bibliography via BibTeX.
%\printbibliography
\bibliography{201115NgwenyaHorowitz01}
\bibliographystyle{apsrev4-2}

\end{document}